\documentclass[3p]{elsarticle}

\usepackage{hyperref}

\journal{Journal of Sound and Vibration}

\biboptions{sort&compress}

\usepackage{amsmath}
\usepackage{color}
\usepackage{overpic}
\usepackage{psfrag}
\usepackage{ulem}
\def\eee {\mathbf{e}}
\def\ff{\mathbf{f}}
\def\uu{\mathbf{u}}

\def\xx{\mathbf{x}}

\def\totalvortvec{\boldsymbol{\omega}}
\def\NoiseInt{K}
\providecommand\bcdot{\boldsymbol{\cdot}}

\providecommand\omegavec{\boldsymbol{\omega}}
\newcommand\Rey{\mbox{\textit{Re}}} 
\newcommand\Sr{\mbox{\textit{Sr}}} 
\newcommand{\tcr}[1]{\textcolor{red}{#1}}
\newcommand{\tcb}[1]{\textcolor{blue}{#1}}

\newcommand{\tcgreen}[1]{\textcolor[rgb]{0,0.6,0}{#1}}

\begin{document}

\begin{frontmatter}

\title{
Processing time-series of randomly forced self-oscillators: \\
the example of beer bottle whistling
} 

\author[caps]{E. Boujo\corref{cor1}} 
\address[caps]{CAPS Laboratory, 
Department of Mechanical and Process Engineering, 
ETH Z\"urich, 
8092 Z\"urich, 
Switzerland}
\ead{eboujo@ethz.ch}

%
\cortext[cor1]{Corresponding authors}


\author[caps]{C. Bourquard}

\author[caps]{Y. Xiong}

\author[caps]{N. Noiray\corref{cor1}}
\ead{noirayn@ethz.ch}

\begin{abstract}
We present a model-based, output-only parameter identification method for self-sustained oscillators forced by dynamic noise, 
which we illustrate  experimentally with a simple aeroacoustic setup:
a turbulent jet impinging a beer bottle
and producing
a distinct whistling tone in a finite range of jet angles and jet velocities.
Given a low-order model of the system, the identification is based on analyzing
stationary time series  
of a single key observable: the acoustic pressure fluctuations inside the bottle.
Noting that this observable exhibits all the characteristics of weakly non-linear self-oscillations,
we choose as a minimal model
the classic Van der Pol (VdP) oscillator:
a linear acoustic oscillator (the bottle) subject to stochastic forcing and non-linear deterministic forcing (the turbulent jet).
Although very simple, the VdP oscillator driven by random noise proves to be a sufficient  
phenomenological description of the aeroacoustic limit cycle for the purpose of model-based linear growth rate identification.
We  derive the associated stochastic amplitude equation, which allows us to describe, on both sides of the  bifurcation, the stochastic fluctuations of the acoustic amplitude  resulting from a competition between 
(i)~linear growth rate induced by the coherent unsteady vortex force, 
(ii)~random forcing induced by the turbulence and
(iii)~non-linear saturation of the coherent flapping motion of the jet. 
Finally, we  use the associated adjoint Fokker-Planck equation to reveal both the deterministic potential well and the stochastic forcing intensity that govern these random fluctuations.
Additional experiments with an external control validate the identification results.
We also observe that different operating conditions can yield similar statistics but different dynamics, 
showing that the identification method can disentangle deterministic and stochastic effects in systems forced  stochastically.
This output-only parameter identification approach can be used in a wide range of phenomena exhibiting stationary self-sustained oscillations  
in the presence of random forcing. 
\end{abstract}


\end{frontmatter}


\section{Introduction}

Self-sustained oscillators constitute an important class of non-linear dynamical systems, with applications in virtually all scientific fields: mechanics, physics, chemistry, biology, sociology, etc.
In the sole case of acoustics, self-oscillations occur in aeroacoustics, thermoacoustics, electroacoustics and vibroacoustics.
Understanding such systems, describing them and predicting their properties are essential to the scientific and technological endeavor, but the presence of measurement noise and dynamic noise often complicates this task.
One possible approach is 
system identification (SI). 
The idea consists in processing measurements to infer either the parameters of a postulated model (parameter identification), or a reduced-order model itself (model identification).

When it is possible to apply a controlled external forcing to the system and to study its response, \textit{input-output SI} is the method of choice.
It generally employs a state-space representation (e.g. autoregressive or moving-average models \cite{LiuShengDing2010}, or finite impulse response models \cite{Polifke2014}) and estimate the parameters by minimizing the prediction error for some output-based quantity, e.g. maximum likelihood \cite{Hamilton94, Shumway11}. 
Recently, Lee \textit{et al.} \cite{Lee2019} applied a white noise of known and varying intensity to a low-density jet and fitted measured probability density functions 
to identify the parameters of a stochastic low-order model of the system.
A suitable model can also be selected among several candidates, rather than postulated, while paying attention to the trade-off between accuracy and simplicity, e.g. with probabilistic Bayesian approaches \cite{Beck2010} or sparsity-promoting algorithms \cite{Chen2014}.

When it is not  possible to apply an external forcing or to measure the input, one must rely on \textit{output-only SI}.
This can be undertaken with a variety of methods, e.g.
modal identification \cite{Nagarajaiah2009}, 
reduced-order modeling \cite{Rowley2017}, 
empirical dynamic modeling \cite{Ye2015},
sparse identification \cite{Brunton2016},
or estimation of the Kramers-Moyal (KM) coefficients \cite{Boettcher06,Friedrich2011}. 
In the latter method, one actually takes advantage of the very presence of dynamic noise to perform output-only SI, because the inherent stochastic forcing drives the system away from its deterministic equilibrium trajectory, thus revealing precious information that would otherwise remain hidden. 
This approach has been applied to analyze stochastic data sets in many systems: 
turbulence \cite{Friedrich97PRL, Renner2002}, 
financial markets \cite{Friedrich2000},
traffic flow \cite{Kriso02}, 
epileptic brain dynamics \cite{Prusseit2007},
earthquakes \cite{Manshour2009},
wind-energy \cite{Milan2013} and 
recently to  thermoacoustic instabilities in combustion chambers \cite{NoirayDenisov16}.
One of the main advantages of the method is the possibility to extract the coefficients of the system's Langevin equation 
or of the corresponding Fokker-Planck equation (FPE).
One practical difficulty lies in so-called finite-time effects that arise, for example, when signals are sampled at a coarse rate, or are band-pass filtered so as to isolate the system's dynamics for a specific range of frequencies of interest. 
An elegant alternative based on the adjoint FPE has been proposed to circumvent this problem and obtain the KM coefficients with improved accuracy \cite{LadePLA09, Honisch11}.
In a previous study \cite{Boujo2017}, we showed how to adapt this method specifically to self-sustained oscillators with weak non-linearities and subject to additive dynamic noise, by reducing the oscillator's second-order stochastic differential equation to a first-order Langevin equation.
We demonstrated the effectiveness of the method with numerical examples.

In the present study, we illustrate this model-based, output-only SI method with a real-word system. We chose this system in the field of aeroacoustics.
Aeroacoustic instabilities can arise when an acoustic element is coupled to a gas flow.
For instance, in Helmholtz resonators \cite{Helmholtz1875, 
Rayleigh1896, 
Cummings1972, 
Cummings1973, 
Panton1975a}
subject to a grazing flow, self-sustained oscillations appear as a  result of a constructive feedback between acoustic oscillations in the resonator and the hydrodynamic response in the neck region \cite{Panton1975b, 
Howe1976, 
Walker1982, 
Khosropour1990, 
Frikha1999, 
MEISSNER2002, 
Ma2009}.
Characterizing aeroacoustic systems, e.g. their linear stability conditions and non-linear properties,  has been the object of extensive research over the past decades. 
While some properties such as aeroacoustic frequencies and oscillation amplitudes can readily be measured, other properties such as linear growth rates are not directly accessible. 
In any case, it is of interest to build models for predicting those properties.
Building accurate predictive models is challenging because aeroacoustic instabilities involve the generation, transport, amplification and saturation of 
vorticity fluctuations in a (possibly turbulent) flow, and strongly depend on the geometry and flow conditions, but significant progress has been made
\cite{Verge1997b, Dequand2003a, Dequand2003b, Auvray2014, Auvray2016}. 
In this context, system-identification is an interesting approach, complementary to direct measurements and predictions.

Specifically, we use in this study the archetypal whistling of a Helmholtz resonator (a beer bottle) submitted to a grazing turbulent jet.
We measure experimentally a single observable of the system, namely the acoustic pressure in the bottle,
 over a range of jet angles and jet velocities.
Considering that the internal acoustic pressure  displays the  signature of weakly non-linear self-oscillations, we start by describing the system phenomenologically as a Van der Pol (VdP) oscillator with weak non-linear damping.
We then derive the stochastic amplitude equation (Langevin equation) that describes the dynamics of the acoustic amplitude in the linearly stable regime (fixed point) and unstable regime (limit cycle), and we obtain the associated Fokker-Planck equation.
Finally, we estimate the KM coefficients and the model parameters with the adjoint FPE method.
To the authors' knowledge, this is the first time output-only identification is performed on an aeroacoustic system by leveraging the presence of noise and estimating the KM coefficients.

The paper is organized as follows.
Section~\ref{sec:Exp} describes the measurements: experimental setup; acoustic power production; and acoustic pressure for whistling and non-whistling conditions.
Section~\ref{sec:physical_model} presents the  amplitude equation that governs the stochastic dynamics of the acoustic amplitude.
In Sec.~\ref{sec:KMcoeffs} the parameters of the low-order model are identified; 
in Sec.~\ref{sec:transient_exp} additional transient experiments validate the identification results; 
in Sec.~\ref{sec:stat_vs_dyn} the stationary and dynamics aspects of the system are discussed in the light of transient simulations. 
Finally, a discussion about the advantages and limitations of the stochastic identification method is given in Sec.~\ref{sec:discussion}.

\section{Experiments}
\label{sec:Exp}

\subsection{Setup}
\label{sec:setup}

The experiment consists in blowing over the neck of a Helmholtz resonator with a turbulent jet.
This setup aims at obtaining various  whistling and non-whistling conditions in a simple and controlled way (results described in Sec.~\ref{sec:Observations}), and at providing measurements that will be used for parameter identification (Sec.~\ref{sec:id}).

\begin{figure} 
\psfrag{y}[][]{\footnotesize $y$ (mm)}
\psfrag{x}[t][]{\footnotesize $x$ (mm)}
\psfrag{xxx}[b][]{\footnotesize $x$ (mm)}
\centerline{
\hspace{-0.2cm}   
   \begin{overpic}[trim=5mm 2mm 0 0mm,clip=true, width=0.35\textwidth,tics=10]{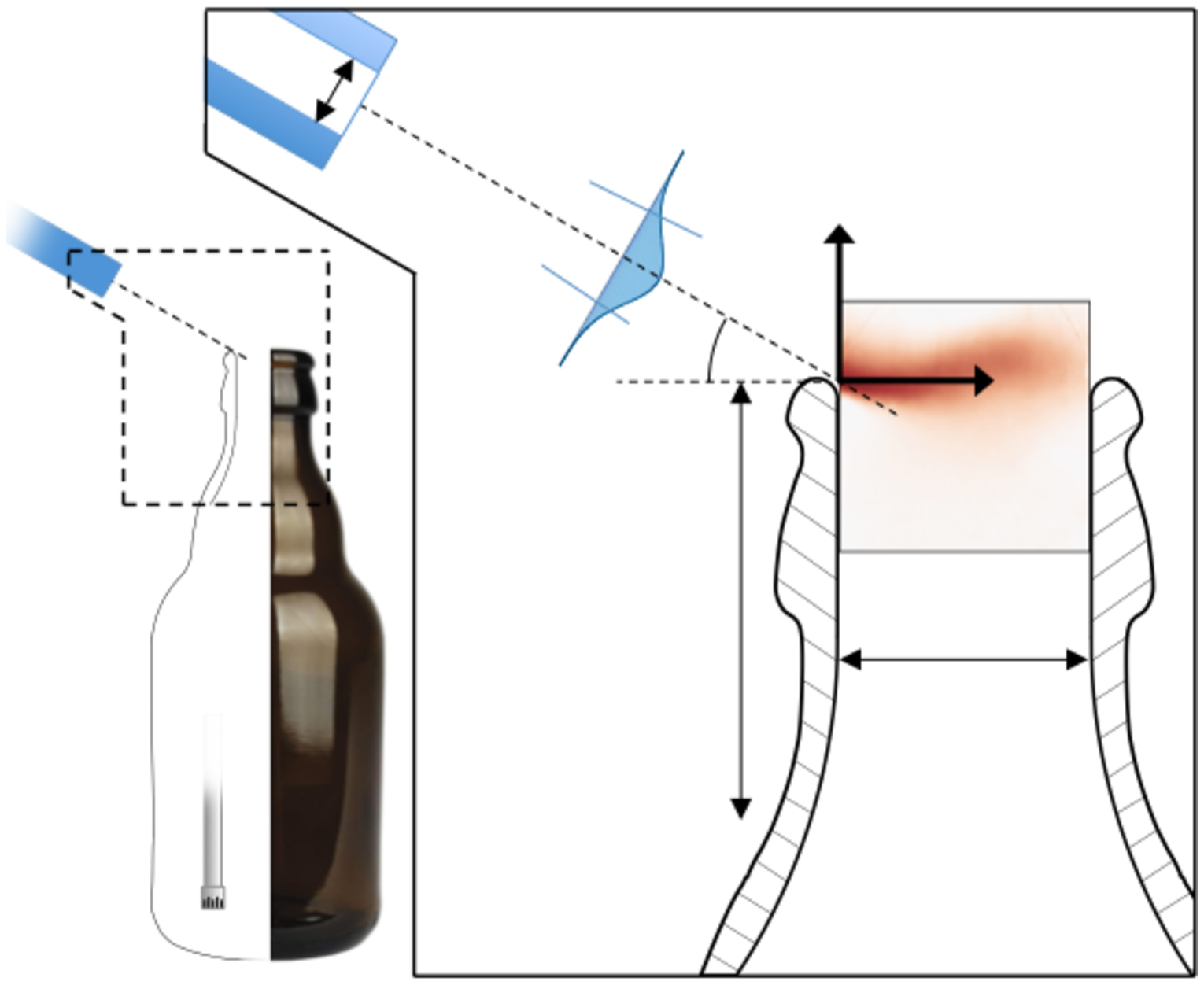} 
      \put(0,80){\footnotesize (a)} 
      \put(20.3,76.7){\footnotesize $d$} 
      \put(32,75){\footnotesize $U$}      
      \put(49.5,69){\footnotesize $\overline{ \boldsymbol{u} }$}      
      \put(54,53){\footnotesize $\theta$} 
      \put(67.5,67.5){\footnotesize $y$} 
      \put(79.5,46.5){\footnotesize $x$} 
      \put(77,29){\footnotesize $D$} 
      \put(55.5,30.5){\footnotesize $L$} 
      \put(68, 3){\footnotesize volume $V$} 
      \put(0,10){\footnotesize micr.} 
   \end{overpic}    
\hspace{1cm}
   \begin{overpic}[trim=0mm 0mm 0mm 0mm,clip, width=0.5\textwidth,tics=10]{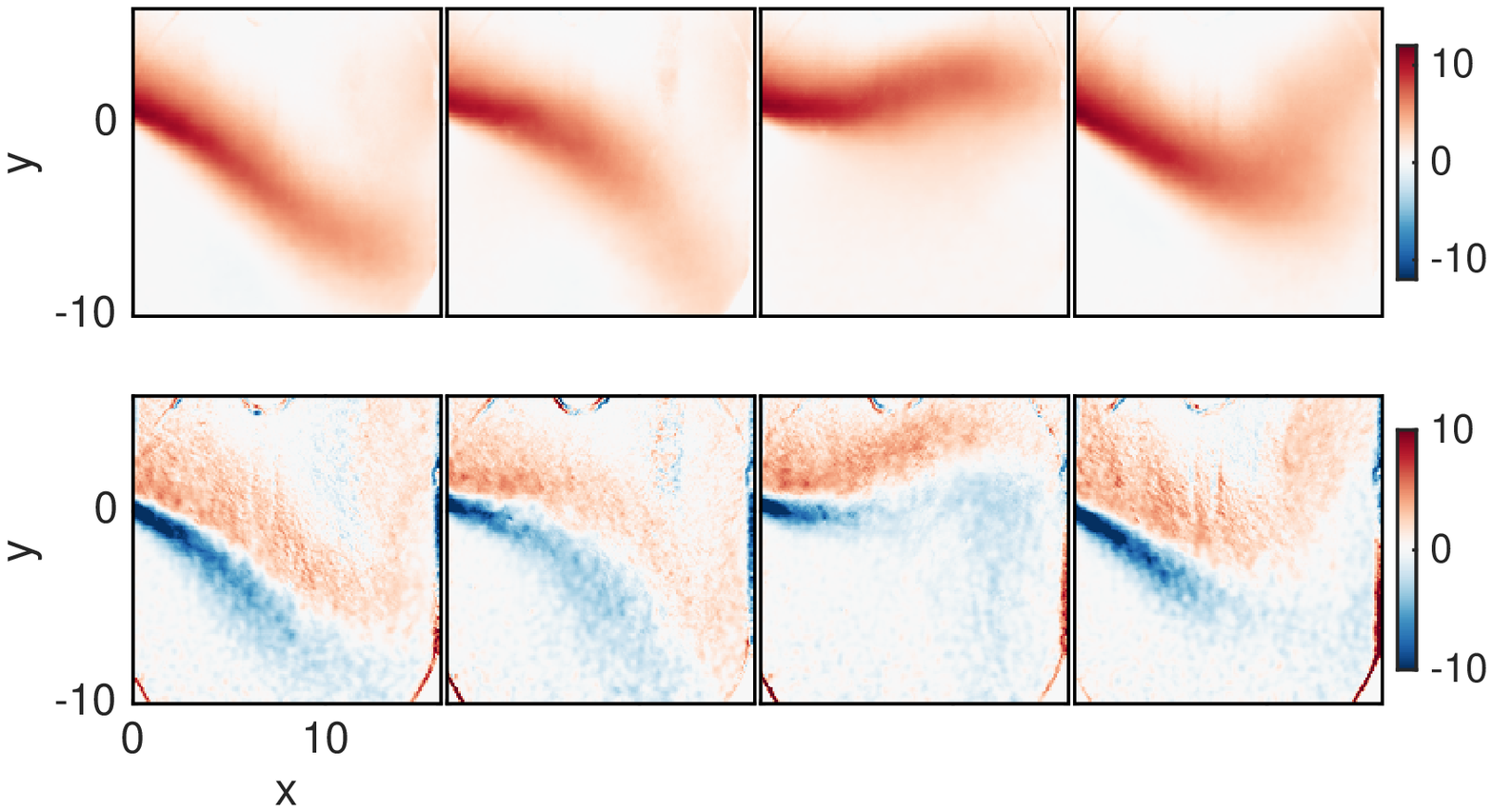}    
      \put(-1,  56){\footnotesize (b)}
      \put(17.4,56){\footnotesize $t_1$} 
      \put(37.8,56){\footnotesize $t_2$} 
      \put(58.2,56){\footnotesize $t_3$} 
      \put(80,  56){\footnotesize $t_4$}
      \put(39,  47){\footnotesize $\uparrow v_{\mathrm{ac}}$}
      \put(81, 38){\footnotesize $\downarrow v_{\mathrm{ac}}$}
      \put(11,36){\footnotesize $\overline u_x + \widetilde u_x$} 
      \put(101,38.8){\rotatebox{90}{\footnotesize (m s$^{-1}$) } }
      \put(11,9.5){\footnotesize $\overline \omega_z +\widetilde \omega_z$}  
      \put(101,8.5){\rotatebox{90}{\footnotesize ($\times 10^3$ s$^{-1}$)}}
   \end{overpic} 
%
\hspace{0.3cm}
\psfrag{xx}[b][]{\footnotesize $x$ (mm)} 
\psfrag{T0/4} [r][r]{\footnotesize $\frac{ T_0}{4}$}
\psfrag{3T0/4}[r][r]{\footnotesize $\frac{3T_0}{4}$}
\psfrag{T0}   [r][r]{\footnotesize        $T_0$}
\psfrag{Int(Int(fx'*ua.dx).dy)}[tr][r]{\footnotesize \tcgreen{$\iint f'_y \,\mathrm{d}x \mathrm{d}y$}, \tcr{$\int \mathcal{P} \,\mathrm{d}x \quad$}} 
\psfrag{Int(fx'*ua.dx)}[][]{\footnotesize $\mathcal{P}(x,t)$}
\psfrag{Int(Int(fx'*ua.dx).dt)}[][][-1][90]{\footnotesize \tcr{$\quad \int \mathcal{P} \,\mathrm{d}t$}}
   \vspace{-0.5cm}
}
\vspace{0.3cm}
\caption{
(a)~Experimental setup. 
(b)~Phase-averaged streamwise velocity and transverse vorticity from PIV measurements ($U$=22~m~s$^{-1}$, $\theta=45^o$). 
(See movies in the supplementary material.)
}
\label{fig:setup}
\end{figure}

The air jet produced by a circular pipe of diameter $d=5$~mm impinges the neck (inner diameter $D=17.2$~mm) of a 350~mL Duvel beer bottle (Fig.~\ref{fig:setup}(a)). 
The pipe has a total length of 3~m ($600 d$).
The jet  mean velocity $U=12-24\pm 0.1$~m~s$^{-1}$ is calculated as $4\dot{m}/ (\bar\rho \pi d^2)$, 
where $\bar\rho$ is the air density and $\dot{m}$ the mass flow measured with a mass flow meter Bronkhorst F-113AC.
The Reynolds number $\Rey = \bar\rho d U / \mu=4000-8000$ (with $\mu$ the air dynamic viscosity) is large enough for the jet to be fully turbulent.
A precision rotary table allows the jet angle to be varied in the range $\theta=0-90 \pm 0.1^o$ around a rotation axis centered on the trailing edge of the upstream rim of the bottle neck.
The distance from the pipe outlet to the rotation axis is $l=37$~mm.
The acoustic pressure in the bottle is recorded with a microphone GRAS 46BD-S2 introduced through the bottle opening via a 3~mm cable shifted away from the 
 plane of symmetry of the tube and bottle system,
and fixed to the inner side of the bottle neck.

Whether whistling occurs or not, depends on the jet velocity $U$ and angle $\theta$.
Figure~\ref{fig:setup}(b) shows, for a whistling case ($U=22$~m~s$^{-1}$, $\theta=45^o$),
measurements of streamwise velocity $u_x=\uu\bcdot\eee_x$ 
and spanwise vorticity $\omega_z=\omegavec\bcdot\eee_z=\partial_x u_y-\partial_y u_x$ 
 obtained in the neck in the  symmetry plane 
 $z=0$ with particle image velocimetry (PIV) 
after phase averaging (keeping time-average fields $\overline u_x$, 
$\overline \omega_z$, 
and coherent fluctuations $\widetilde u_x$, 
$\widetilde \omega_z$, while removing incoherent fluctuations) 
at different phases of the acoustic cycle. 
PIV is performed by seeding the jet with DEHS particles (mean diameter 1~$\mu$m), 
illuminating the  
plane with a 0.5~mm laser sheet (double-pulse, 532~nm, 2$\times$6~mJ at 10~kHz), and collecting the particles Mie scattering  with a high-speed CMOS camera.
These fields illustrate the jet structure and its transverse motion at the whistling frequency.

\begin{figure} 
\centerline{
   \hspace{-0.3cm}
   \psfrag{x}[t][]{\small $x$ (mm)}
   \psfrag{phase}[r][][-1][90]{\small $t$ (s)}
   \psfrag{T0/3} [r][r]{\small $\frac{ T_0}{3}$}
   \psfrag{2T0/3}[r][r]{\small $\frac{2T_0}{3}$}
   \psfrag{T0}   [r][r]{\small        $T_0$}
   \psfrag{Int(fx'.dx)}[t][]{\small \tcgreen{$\int f'_y \,\mathrm{d}y$}} 
   \begin{overpic}[trim=0mm -37mm 0mm 0mm,clip, width=0.385\textwidth,tics=10]{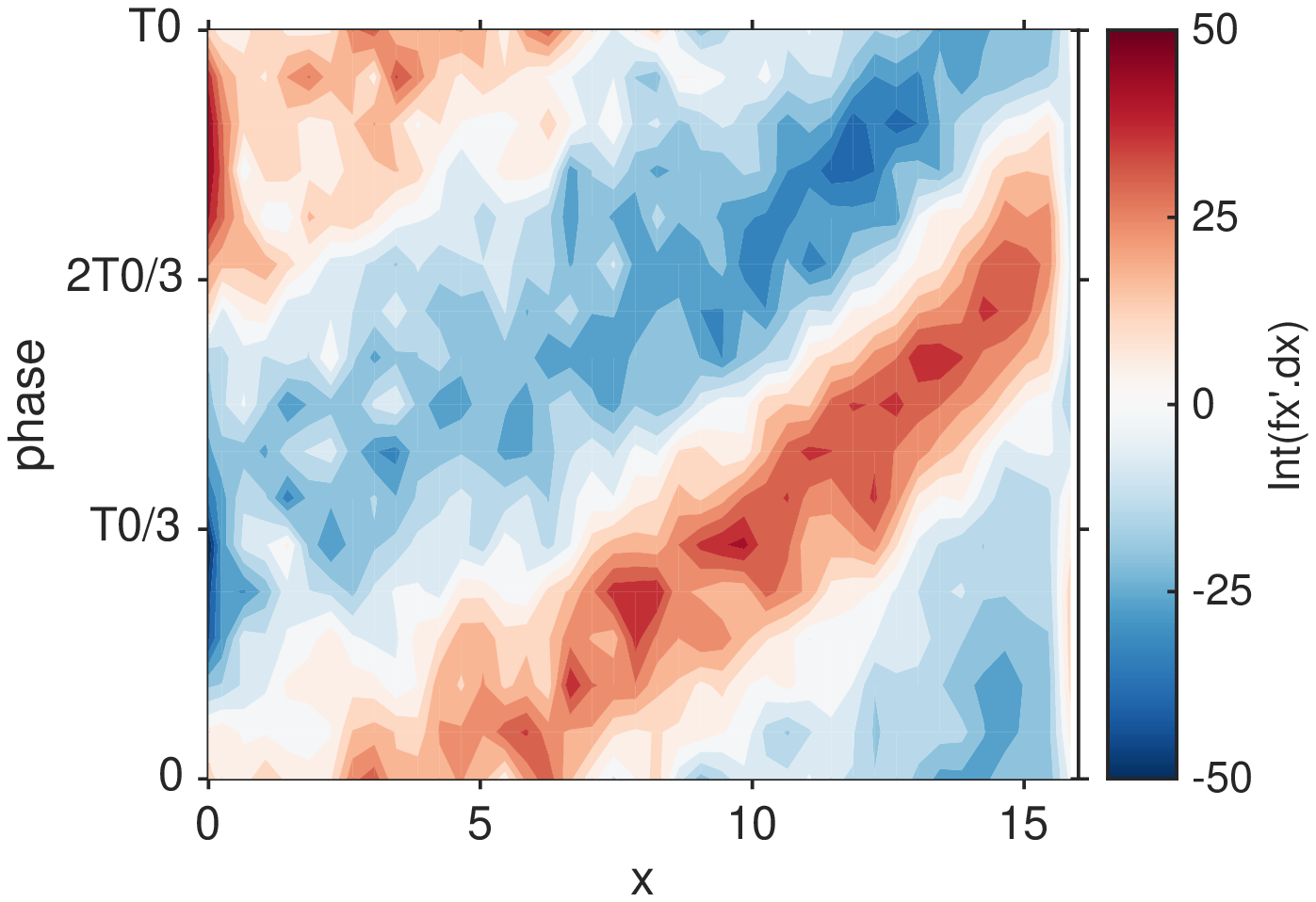}
      \put(-11,92){\footnotesize (a)} 
   \end{overpic} 
   \hspace{1cm}
   \psfrag{xx}[b][]{\footnotesize $x$ (mm)} 
   \psfrag{phase}[t][]{\footnotesize $t$ (s)}
   \psfrag{T0/4} [r][r]{\footnotesize $\frac{ T_0}{4}$}
   \psfrag{3T0/4}[r][r]{\footnotesize $\frac{3T_0}{4}$}
   \psfrag{T0}   [r][r]{\footnotesize        $T_0$}   
   \psfrag{00}   [r][r]{\footnotesize        $0$}
%
   %
   %
%
   \begin{overpic}[trim=0mm 5mm 0mm 0mm,clip, width=0.41\textwidth,tics=10]{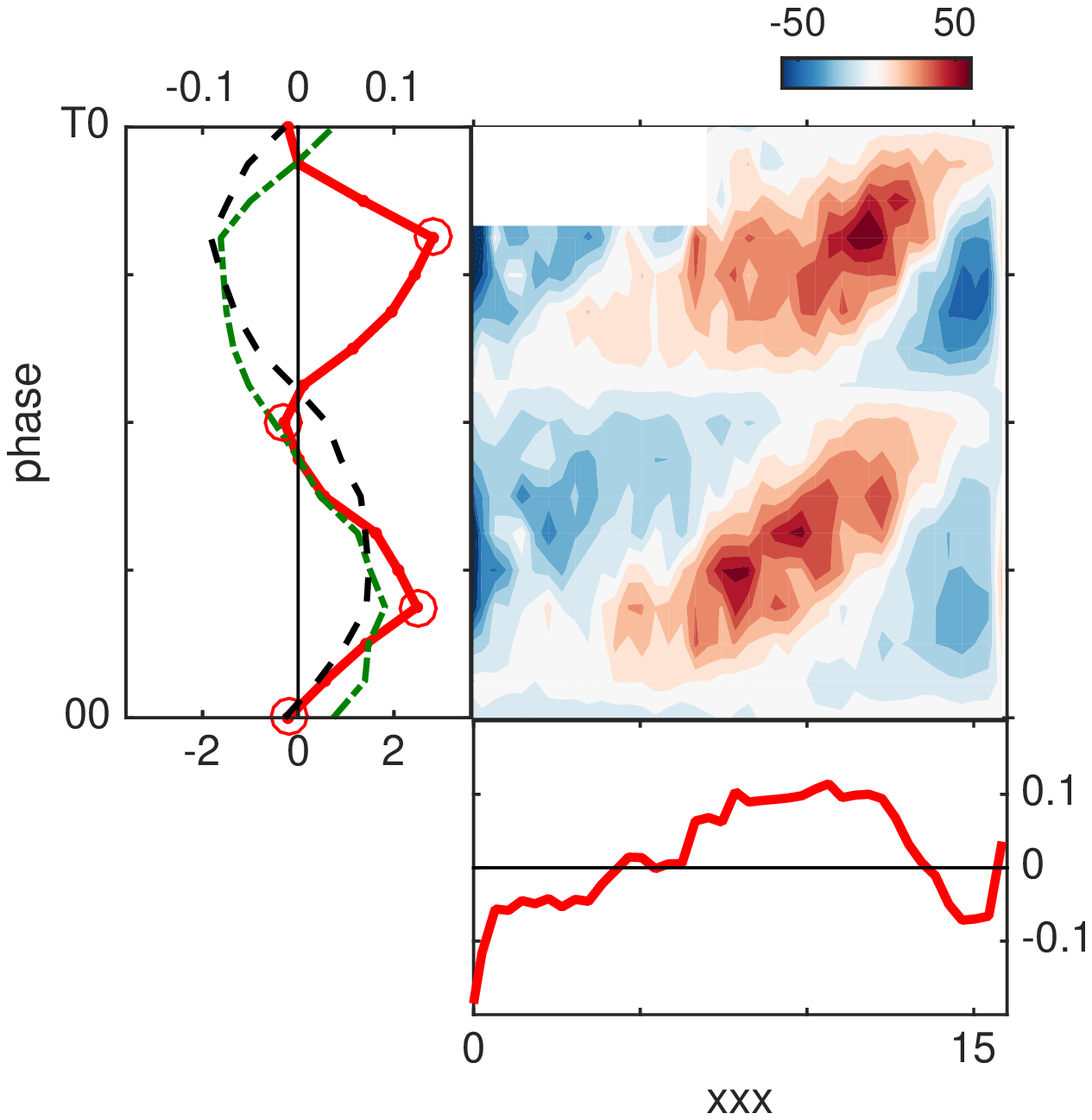}  
      \put(-6,86){\footnotesize (b)} 
      \put(11,97){\footnotesize \tcgreen{$\iint f'_y \,\mathrm{d}\xx$}, \tcr{$\int \mathcal{P} \,\mathrm{d}x \quad$}} 
      \put(32,75){\footnotesize $t_4$} 
      \put(21,55){\footnotesize $t_3$} 
      \put(38,48){\footnotesize $t_2$} 
      \put(21,36){\footnotesize $t_1$} 
      \put(16,23){\footnotesize $v_{\mathrm{ac}}$ (m~s$^{-1}$)}
      \put(46.5,82){\footnotesize $\mathcal{P}(x,t)$}
      \put(97,19){\footnotesize \tcr{$\quad \int \mathcal{P} \,\mathrm{d}t$}}
      \put(59,-2){\footnotesize $x$ (mm)}
   \end{overpic}
   \vspace{-0.3cm}
}
\vspace{0.3cm}
\caption{
(a)~ Space-time evolution of $\int f'_y\,\mathrm{d}y$ 
(in N~m$^{-2}$), the vertical component of the fluctuating vortex force integrated vertically at each streamwise location. 
This is the force per unit neck-cross-section area in the  symmetry plane.
(b)~Acoustic power density $\mathcal{P}$ 
(in N~m$^{-1}$s$^{-1}$)
as a function of streamwise location and time;
temporal and vertical integrals of $\mathcal{P}$ (in N~m$^{-1}$ and N~s$^{-1}$ respectively); 
time evolution of the acoustic velocity $v_{\mathrm{ac}}$  and space-integrated fluctuating vertical force $f'_y$ (in N~m$^{-1}$). 
}
\label{fig:P}
\end{figure}

\begin{figure} 
\psfrag{x}[t][]{\small $x$ (mm)}
\psfrag{y}[t][]{\small $y$ (mm)}
\psfrag{xx}[t][]{\small $x$ (mm)}
\psfrag{yy}[b][]{\small $y$ (mm)}
\psfrag{vac}[][]{\small $v_{\mathrm{ac}}$ (m~s$^{-1}$)}
\psfrag{uac}[][]{\small $u_{\mathrm{ac}}$ (m~s$^{-1}$)}
\centerline{
   \begin{overpic}[trim=0mm 0mm 0mm 0mm, clip=true, height=5cm,tics=10]{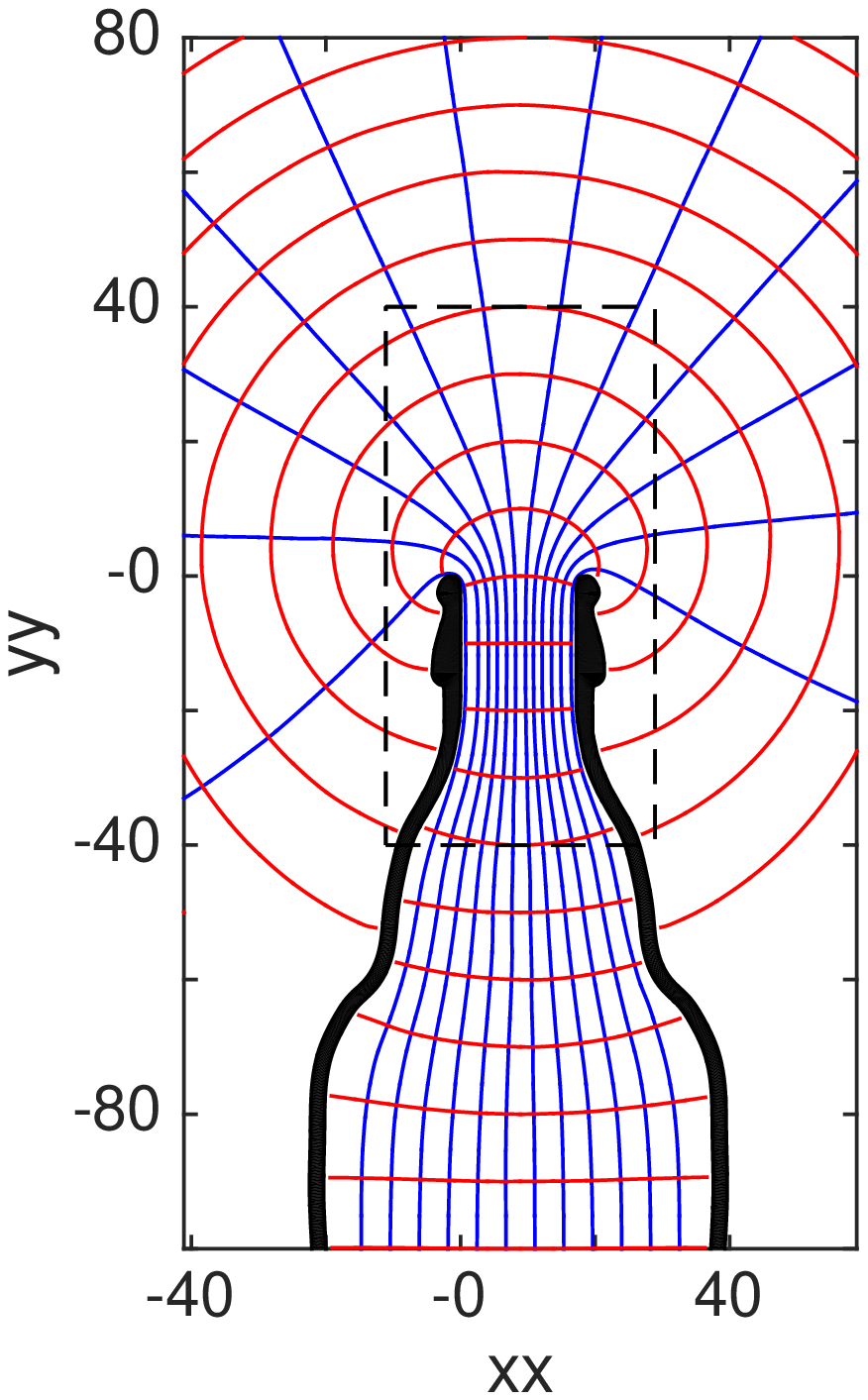}
      \put(-4,95){\footnotesize (a)} 
   \end{overpic} 
   \hspace{-0.1cm}
   %
   \begin{overpic}[trim=0 0 0 0, clip=true, height=5cm,tics=10]{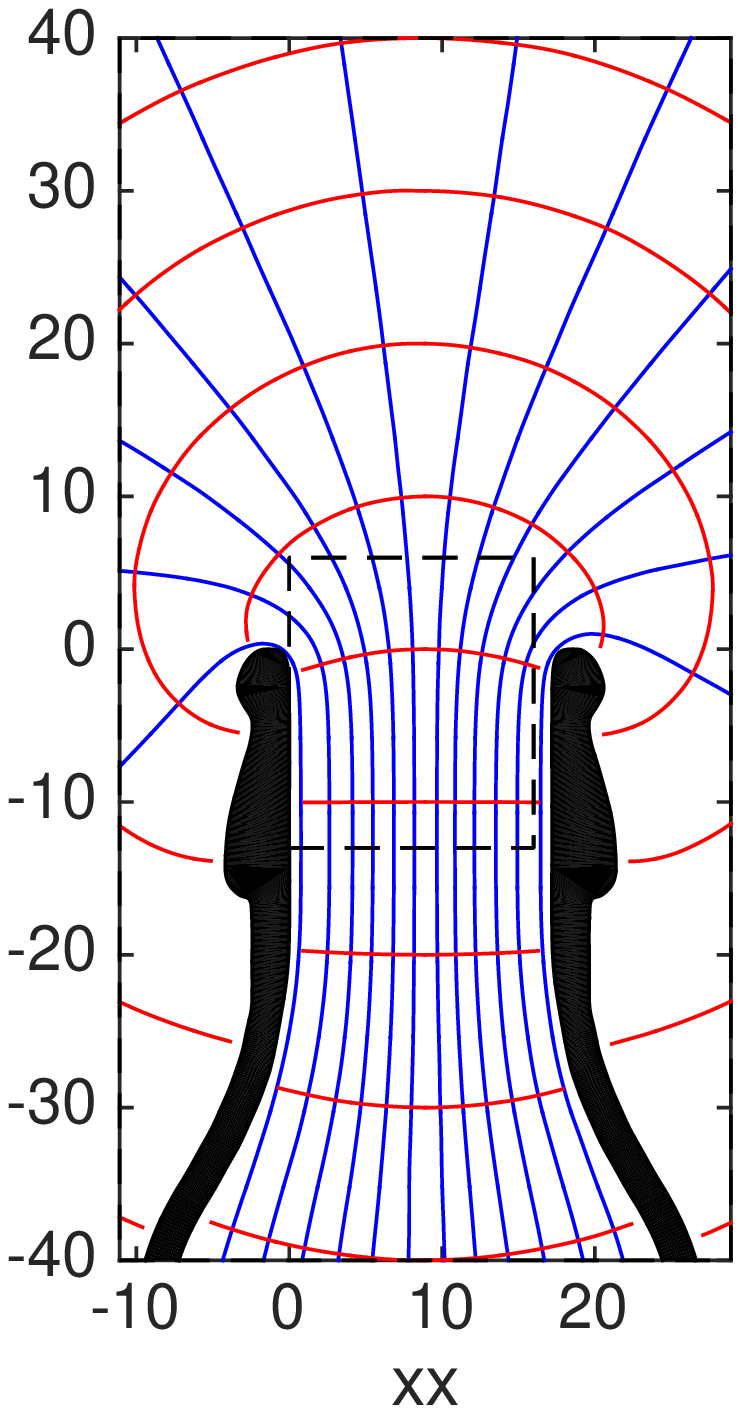}
   \end{overpic} 
   \hspace{-0.05cm}
   %
   \begin{overpic}[trim=2mm 0mm 0mm 0mm, clip=true, height=4.92cm,tics=10]{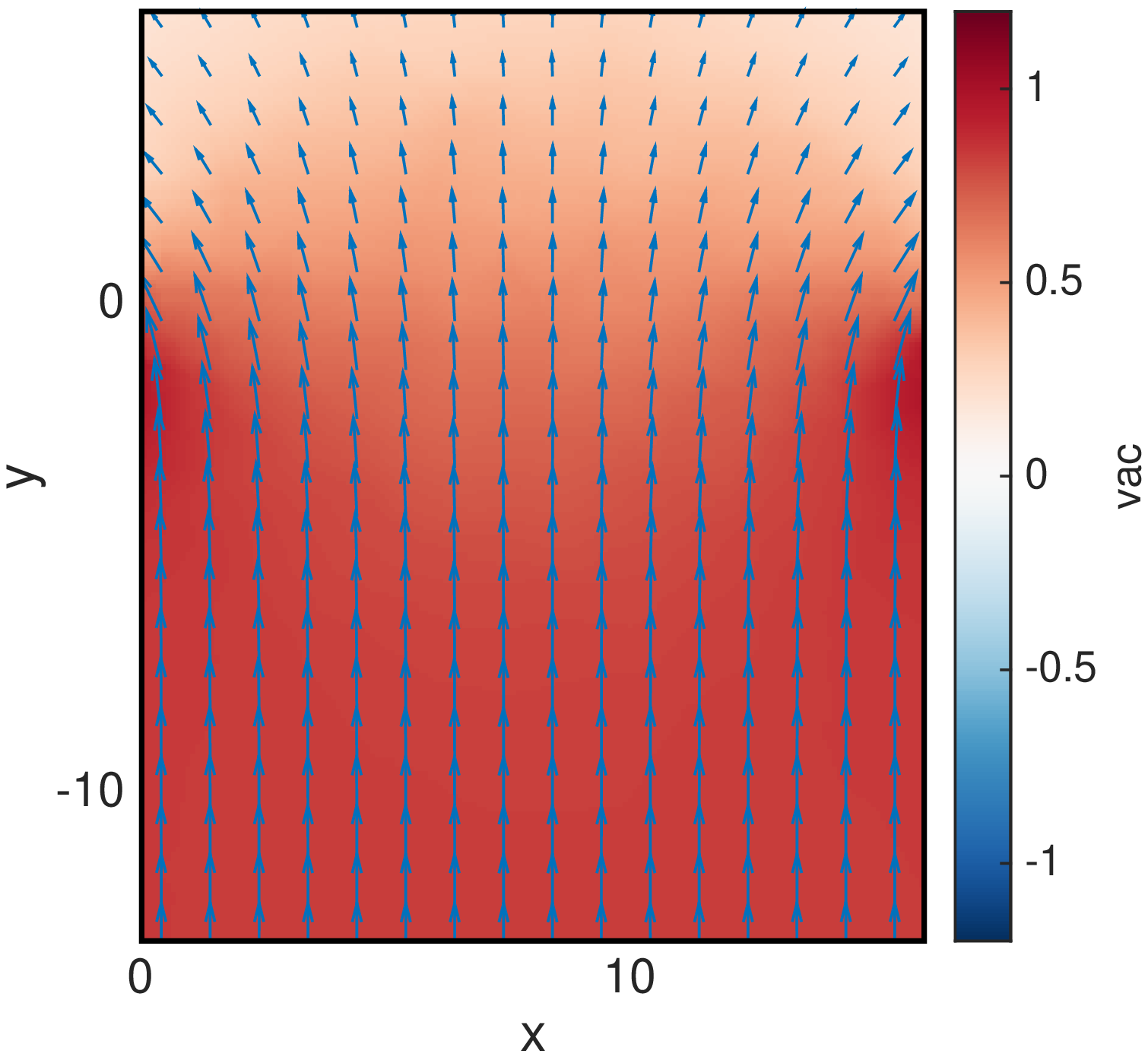}
      \put(1,89){\footnotesize (b)} 
   \end{overpic} 
   \begin{overpic}[trim=2mm 0mm 0mm 0mm, clip=true, height=4.92cm,tics=10]{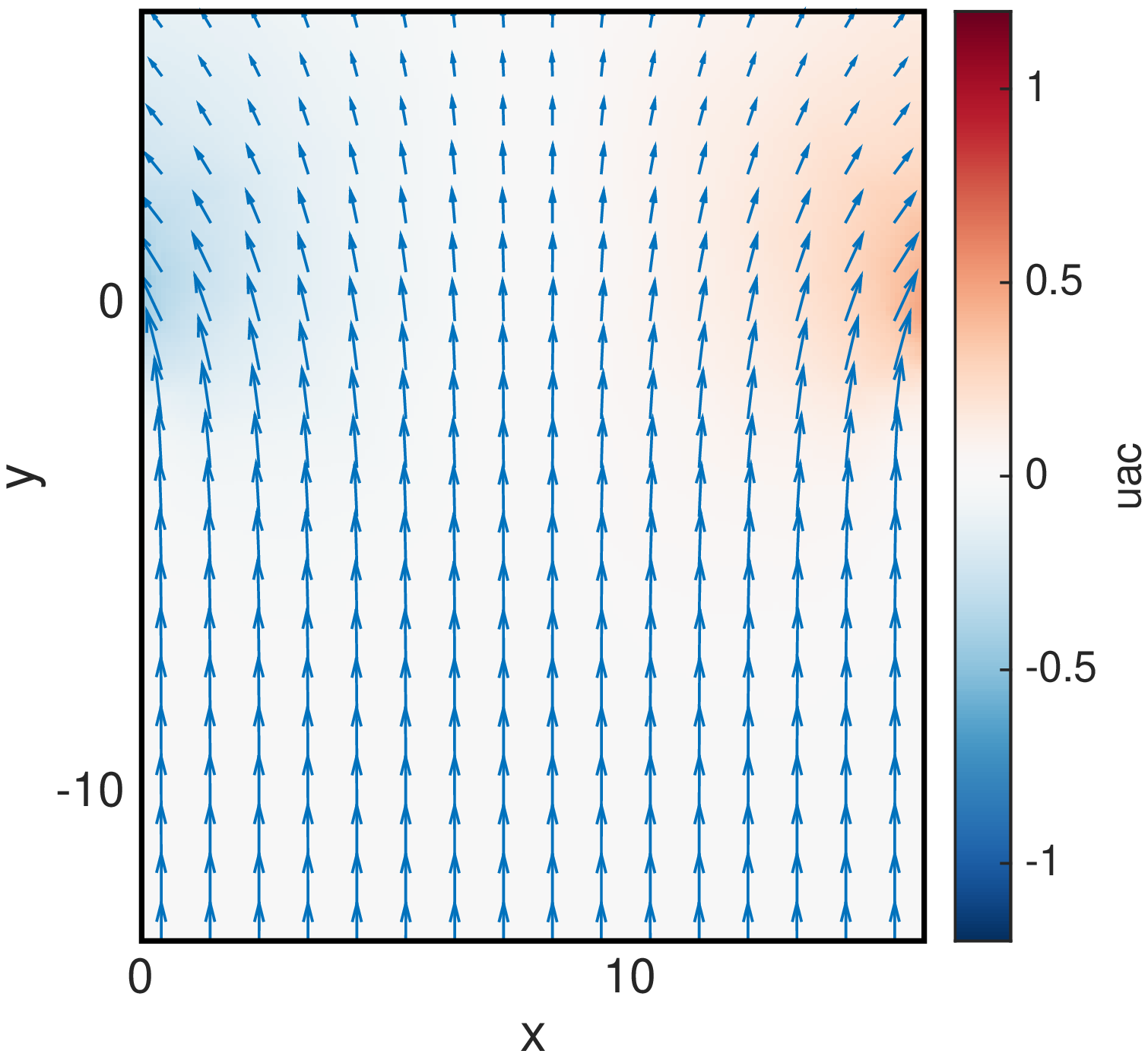}
   \end{overpic} 
}
\caption{
Dominant acoustic mode computed with a linear Helmholtz solver.
(a)~Equipotentials (red) and streamlines (blue) in the symmetry plane $z=0$,
showing the structure of a compact monopole.
(b)~Vertical and horizontal velocity fields in the same plane.
The linear mode is normalized such that the largest acoustic velocity magnitude is 1.
}
\label{fig:avsp}
\end{figure}

\subsection{Acoustic power production and dissipation}
\label{sec:ac_power}

The bottle and the jet constitute an aeroacoustic system, where coupling occurs through mutual interaction: the turbulent jet  acts as a forcing upon the acoustic field of the Helmholtz resonator, and acoustic fluctuations in the bottle neck exert a feedback forcing on the jet.
Thus, whistling is a self-sustained aeroacoustic oscillation, where part of the 
kinetic energy from the turbulent jet is pumped into the acoustic field. 
It is possible to leverage this PIV data further by considering Howe's 
energy corollary
 \cite{Howe80}, which expresses how the fluctuating component of the Lamb vector $\totalvortvec\times\uu$ induces a vortex force 
\begin{equation}
\ff'(x,y,t)= \bar\rho (\totalvortvec\times\uu)'
\end{equation}
that does work on the acoustic field
(throughout the paper, primes denote fluctuating quantities, that need not be small).
Figure~\ref{fig:P}(a) shows the  vertical component of the force
\begin{eqnarray} 
f'_y
=
\ff'\bcdot\eee_y 
= 
- \bar\rho (\totalvortvec \times \uu)'\bcdot\eee_y
\simeq - \bar\rho (\omega_z u_x)',
\end{eqnarray}
measured in the symmetry plane and
integrated vertically across the jet along $y$.
The force exhibits a clear wave structure  in the streamwise direction $x$ (with one wavelength across the neck), and spatio-temporal propagation downstream at  $U_p \simeq 3$~{m~s$^{-1}$}.
This propagation velocity, substantially smaller than the jet velocity at the pipe outlet (22~{m~s$^{-1}$}),
is consistent with a typical convection velocity for hydrodynamic perturbations of approximately 0.4 times the local velocity of the  turbulent jet.
The local jet velocity $U_j$ can be estimated from the PIV measurements as well as from the analytical expression 
$U_j(x_p) = U_j(0)/(1+x_p/(4.5d))$,  with 
$x_p$ the propagation distance measured from the pipe outlet,
and $U_j(0)=U$ the jet mean velocity in the pipe \cite{Pope00}.
For $U=22$~{m~s$^{-1}$}, this yields
$U_j \in [6.5, \, 8.3]$~m~s$^{-1}$ in the bottle neck (downstream and upstream rims at 
$x_p = x_d \simeq l + D = 54$~mm and 
$x_p = x_u =      l     = 37$~mm, respectively);
therefore the propagation velocity in the neck is of the order of
$U_p \simeq 0.4 U_j \in [2.6, \, 3.3]$~m~s$^{-1}$.
The spatial distribution of the acoustic velocity field $\uu_{\mathrm{ac}}$ which is used for the evaluation of $\mathcal{P}(x,t)$ is computed using a Helmholtz solver (see Fig.~\ref{fig:avsp}), and the temporal evolution is adjusted based on the PIV data at the bottom left corner of the field of view which is vorticity-free all along the acoustic cycle.

Figure~\ref{fig:P}(b) shows 
the acoustic power density 
given to or taken from the acoustic field across the neck during an oscillation cycle  \cite{Howe80}, 
\begin{align} 
\mathcal{P}(x,t) 
=  \int \ff' \bcdot \uu_{\mathrm{ac}} \,\mathrm{d}y 
= - \int \bar\rho (\totalvortvec \times \uu)' \bcdot \uu_{\mathrm{ac}} \,\mathrm{d}y.
\label{eq:Pac1}
\end{align}
Importantly, the aeroacoustic coupling inside the neck is three dimensional, so the map generated from the  symmetry plane velocity field only gives qualitative information. 
Also, it does not reflect the production from the unstable base flow (steady solution of the incompressible Navier--Stokes equations), because $\mathcal{P}$ is processed from limit-cycle data, where the effective gain from the non-linear aeroacoustic feedback balances the linear acoustic damping (visco-thermal and radiation losses). 
That said, it is enlightening to observe that, for this limit cycle ($U$=22~m~s$^{-1}$, $\theta=45^o$), regions near the upstream and downstream rims are, on average, dissipating acoustic energy, while the central region is producing acoustic energy, as shown by the power density integrated over one acoustic period, denoted $\int \mathcal{P}\,\mathrm{d}t$.
The instantaneous net effect over the
neck width, 
$\int \mathcal{P}\,\mathrm{d}x$, 
is positive at all times. 
This acoustic power production is the result of a good synchronization between the vertical acoustic velocity $ v_{{\mathrm{ac}}} = \uu_{\mathrm{ac}}\bcdot\eee_y$ 
and the vertical force $f_y'=\ff'\bcdot\eee_y$ 
(approximately equal to $-\bar\rho (\omega_z u_x)'$ given the system geometry), yielding production  when $v_{{\mathrm{ac}}}$ is directed both outward and inward.

While the above  observations are interesting to characterize qualitatively the production and dissipation of acoustic power, we stress that our phenomenological model  (Sec.~\ref{sec:physical_model}) is a description of the system based on the  acoustic pressure inside the bottle,  and that the parameter identification (Sec.~\ref{sec:id}) only requires measurements of this observable.

\begin{figure}
 \psfrag{eta}[l][r]{}
\psfrag{A}[l][][1][-90]{\footnotesize $A$ (mbar)}
 \psfrag{etadot/wn}[t][]{\footnotesize $\dot p/\omega_a$ (mbar)}
\psfrag{P(eta)}[][]{}
\psfrag{t}[t][]{\footnotesize $t$ (s)}
\psfrag{P(A)}[t][]{\footnotesize $\,\,\,P_\infty(A)$}
\centerline{   
   \hspace{-0.1cm}
   \begin{overpic}[width=0.65\textwidth, clip=true, tics=10]{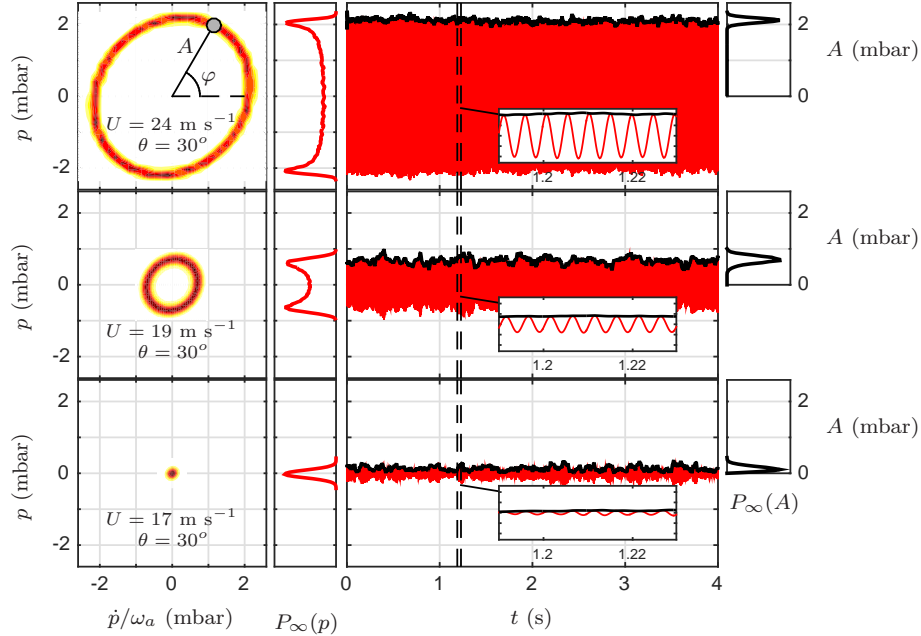}
      \put(-1.6,59.5){\rotatebox{90}{\footnotesize $p$ (mbar)} }
      \put(-1.6,36. ){\rotatebox{90}{\footnotesize $p$ (mbar)} }
      \put(-1.6,12.5){\rotatebox{90}{\footnotesize $p$ (mbar)} }
      \put(10.,  60.9){\scriptsize $U=24$~m~s$^{-1}$}
      \put(13.9, 58.3){\scriptsize $\theta=30^o$}
      \put(10.,  35.2){\scriptsize $U=19$~m~s$^{-1}$}
      \put(13.9, 32.6){\scriptsize $\theta=30^o$}
      \put(10.,  11.8){\scriptsize $U=17$~m~s$^{-1}$}
      \put(13.9,  9.2){\scriptsize $\theta=30^o$}
      \put(18.7, 70.4){\footnotesize $A$}
      \put(21.7, 66.9){\footnotesize $\varphi$}
      \put(27,-1){\footnotesize $\quad\,\, P_\infty(p)$}
   \end{overpic}  }
\caption{
Sample acoustic pressure signal $p(t)$ inside the bottle (red), 
corresponding envelope $A(t)$ (black),
 their stationary PDFs $P_\infty(p)$ and $P_\infty(A)$,
and the stationary joint PDF $P_\infty(p, \dot p/\omega_a)$
for a non-whistling condition ($U=17$~m~s$^{-1}$) and two whistling conditions ($U=19$ and 24~m~s$^{-1}$). 
}  
\label{fig:eta_A_signals_PDFs}
\end{figure}

\begin{figure}
\psfrag{Umean [m/s]}[t][]{\footnotesize $U$ (m~s$^{-1}$)}
\psfrag{U}[t][]{\footnotesize $U$ (m~s$^{-1}$)}
\psfrag{Sr}[t][]{\footnotesize \Sr}
\psfrag{vac/U(xu)}[b][]{\footnotesize $v_{\mathrm{ac,rms}}/U_j(x_u)$}
\psfrag{std(eta) [mbar]}[b][]{\footnotesize $p_\text{rms}$ (mbar)}
\psfrag{std(eta) [mbar] v2}[t][]{\footnotesize $p_\text{rms}$ (mbar)}
\psfrag{stdeta2}[b][]{\footnotesize $p_\text{rms}^2$ (mbar$^2$)}
\psfrag{theta [deg]}[b][]{\footnotesize $\theta$ $(^o)$}
\psfrag{th=40dg}[l][l]{\scriptsize \,$\theta=40^o$}
\psfrag{th=35}  [l][l]{\scriptsize \,$\theta=35^o$}
\psfrag{th=30}  [l][l]{\scriptsize \,$\theta=30^o$}
\psfrag{th=25}  [l][l]{\scriptsize \,$\theta=25^o$}
%
\centerline{   
   \hspace{0.3cm}
   \begin{overpic}[trim=0 0 0 0, clip, width=0.3\textwidth,tics=10]{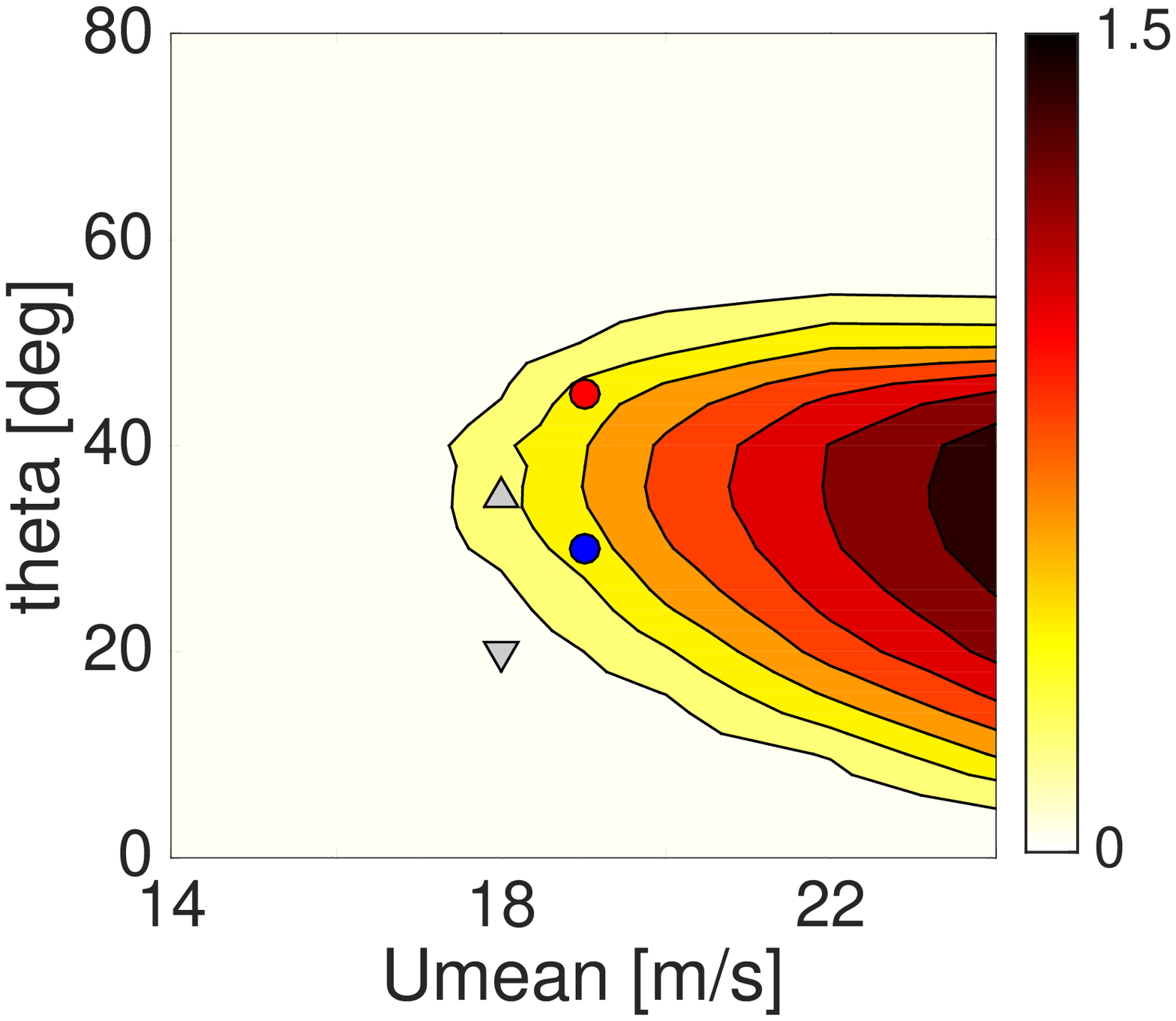}
      \put(-5,80){\footnotesize (a)}
      \put(17,77){\footnotesize $p_\text{rms}$ (mbar)}
   \end{overpic}      
   \hspace{0.2cm}
   \begin{overpic}[width=0.318\textwidth,tics=10]{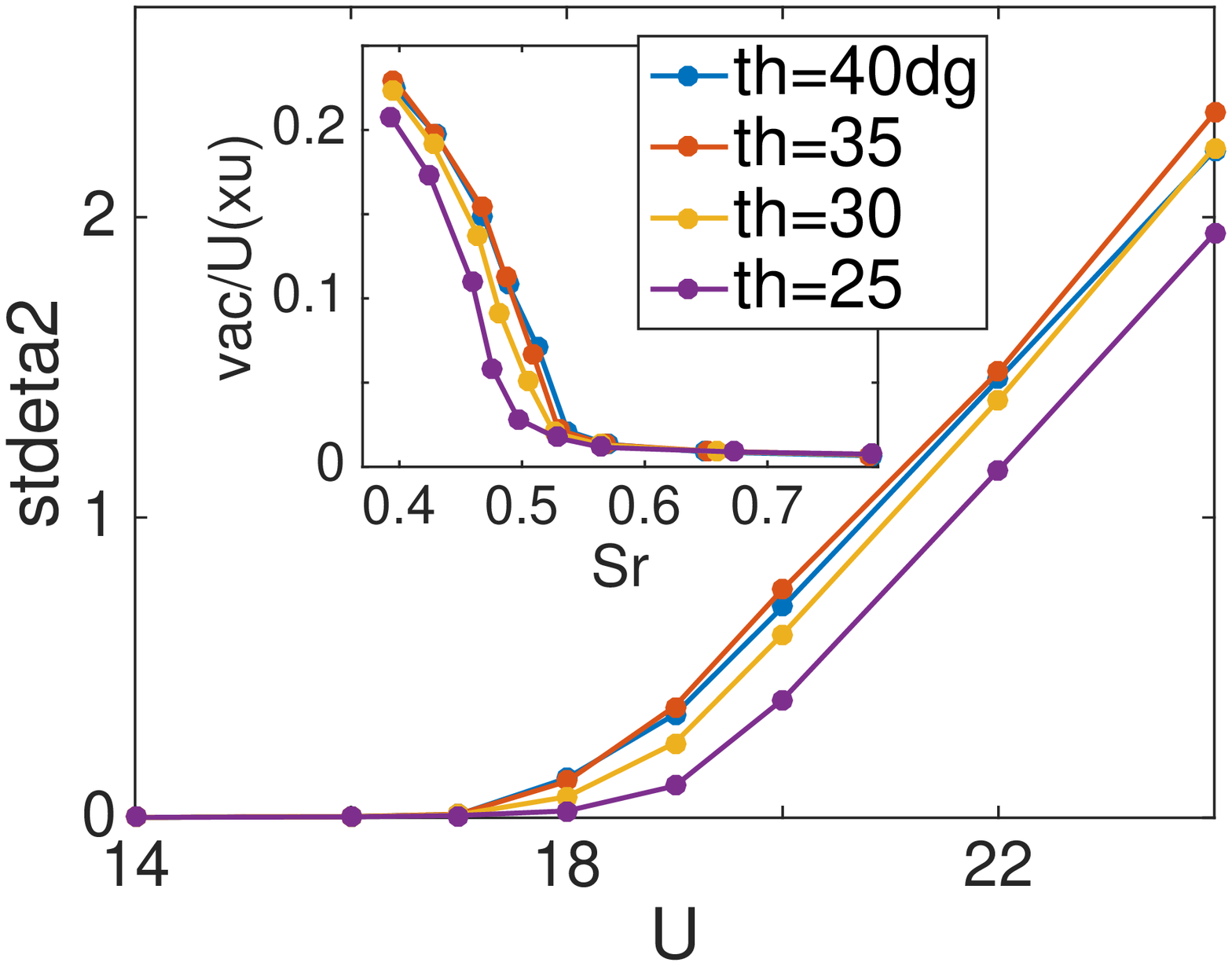}     
      \put(-5,75){\footnotesize (b)}
   \end{overpic}   
   \psfrag{th [deg]}[r][][1][-90]{\footnotesize $\theta$ $(^o)$}
   \hspace{0.3cm}
   \begin{overpic}[trim=0 0 0 0, clip, width=0.302\textwidth,tics=10]{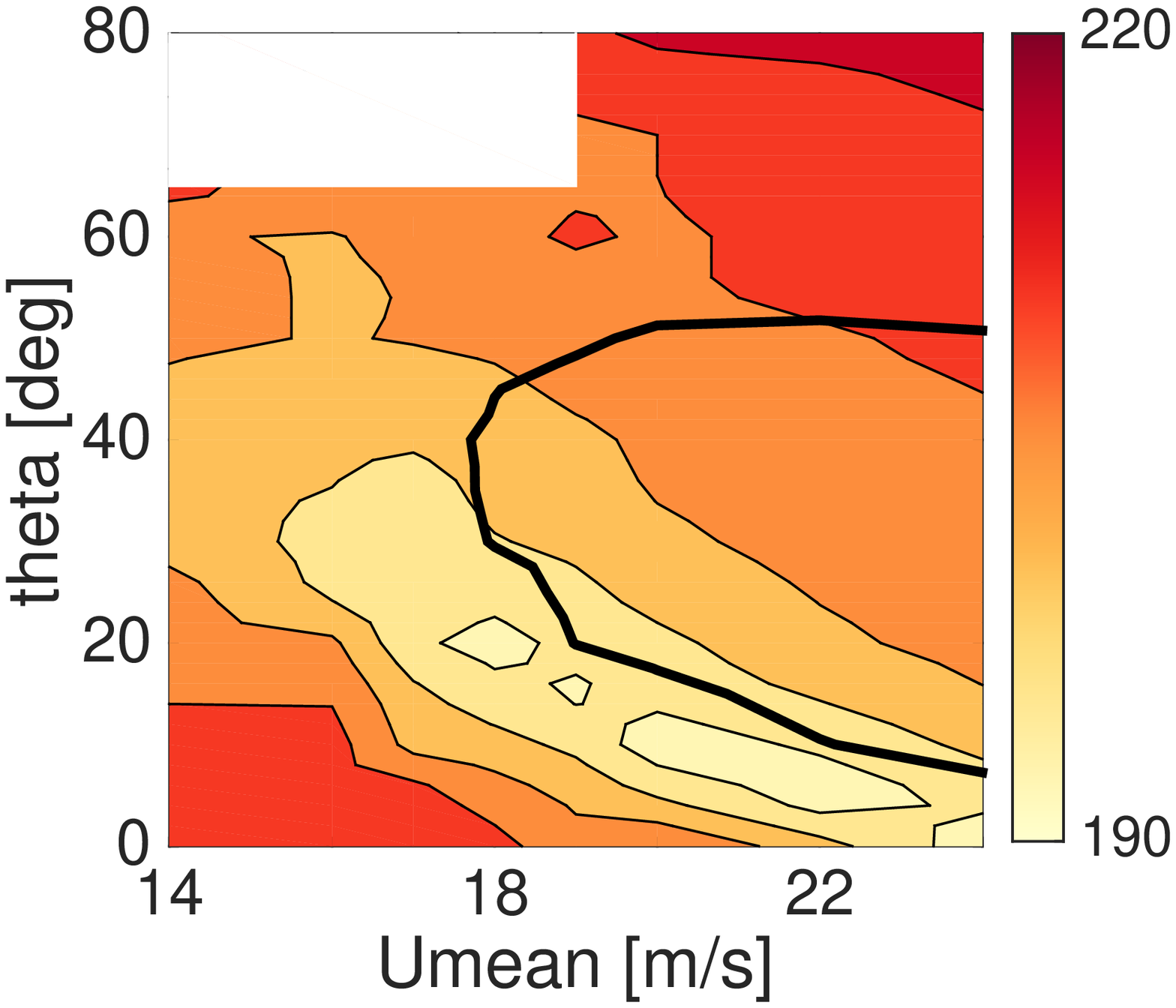}
       \put(-5,79){\footnotesize (c)}
       \put(17,75){\footnotesize $\omega_a/2\pi$ (Hz)}
   \end{overpic} 
}
\vspace{0.5cm}
\centerline{  
\psfrag{f}              [t][]{\footnotesize $\omega/2\pi$ (Hz)}
\psfrag{10log|PSDp|}    [][]{\footnotesize  $20\log_{10} \left( \frac{p_{\mathrm{rms}} }{ p_{\mathrm{ref}} } \right)$  (dB)}
\psfrag{U=17 m/sec}[lt][lt]{\scriptsize $U=17$~m~s$^{-1}$}
\psfrag{U=19 m/sec}[lt][lt]{\scriptsize $U=19$~m~s$^{-1}$}   
\psfrag{U=24 m/sec}[lt][lt]{\scriptsize $U=24$~m~s$^{-1}$}
   \begin{overpic}[trim=0 0 0 0mm, clip=true, width=0.9\textwidth,tics=10]{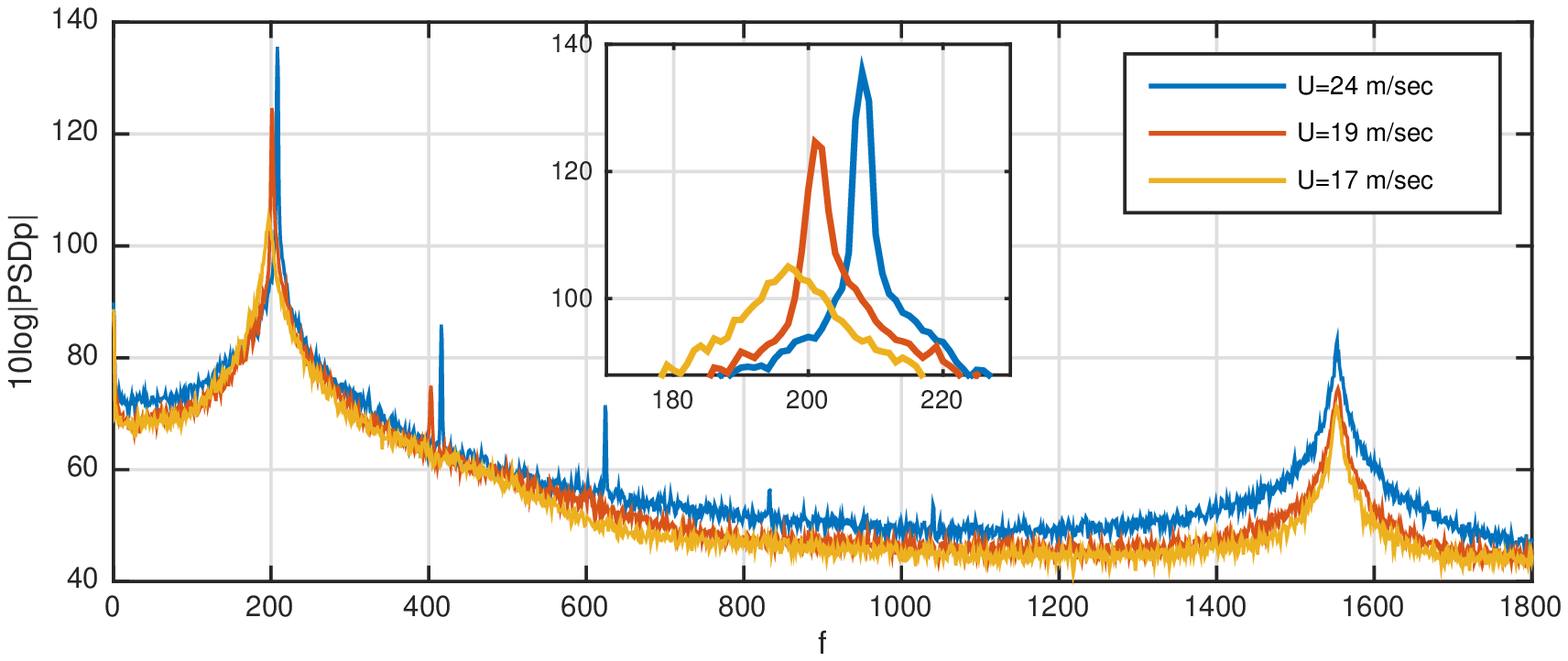}
       \put(-1,40){\footnotesize (d)}
   \end{overpic}  
}
\caption{
(a)~Acoustic pressure fluctuations $p_\text{rms}$  vs. jet velocity $U$ and jet angle $\theta$.
Triangles show conditions of Fig.~\ref{fig:controlONOFF},
circles show conditions of 
Fig.~\ref{fig:same_pdf_different_dynamics}.
(b)~Scaling $p_\text{rms}^2 \propto U-U_c$ above the onset of whistling at a critical velocity $U_c$.
Inset: same data shown in dimensionless form (acoustic velocity oscillation amplitude in the bottle neck normalized by the jet velocity at the upstream rim, against Strouhal number).
(c)~Aeroacoustic frequency vs. $U$ and $\theta$ (thick line: stability limit $\nu=0$; see Fig.~\ref{fig:coeffs}).
(d)~Frequency spectra for jet velocities between $U$=17 and 24~m~s$^{-1}$ ($\theta=30^o$).
}  
\label{fig:prms_wa_spectra}
\end{figure}

\subsection{Acoustic pressure}
\label{sec:Observations}

Varying the jet velocity and angle leads to distinct behaviors.
Figure~\ref{fig:eta_A_signals_PDFs} shows in red time signals of acoustic pressure $p(t)$  recorded for the angle $\theta=30^o$ and the three velocities 
$U=17,$ 19 and 24~{m~s$^{-1}$}.
Since those signals are quasi-harmonic (see insets),
their slowly varying envelope $A(t)$ can be extracted with the Hilbert transform, as shown in black.
For $U$=19 and 24~{m~s$^{-1}$} (intense whistling), the signal is characterized by large-amplitude harmonic oscillations at 
a frequency $f_a = \omega_a/(2\pi)$ that we denote the aeroacoustic frequency, 
and the random fluctuations of the envelope result from the stochastic forcing of the turbulent jet.  
In contrast,
for $U$=17~{m~s$^{-1}$} (no whistling), the limit cycle disappears and one retrieves the dynamics of a noise-driven linearly stable oscillator. 

It is interesting to look at the probability density functions (PDFs) $P_\infty(p)$ and $P_\infty(A)$ in the stationary regime, defined by the probabilities 
\begin{align}
\int_{p_1}^{p_2} P_\infty(p) \, \mathrm{d}p
\quad \mbox{and } \quad
\int_{A_1}^{A_2} P_\infty(A) \, \mathrm{d}A
\end{align}
of the signal falling within an interval $p(t)\in[p_1,p_2]$ 
 and $A(t)\in[A_1,A_2]$,
respectively.
In practice, these PDFs can be obtained by binning the signals (i.e. constructing histograms), and normalizing such that 
$\int_{-\infty}^\infty P_\infty(p) \,\mathrm{d}p = 1$
and
$\int_0^\infty P_\infty(A) \,\mathrm{d}A = 1$.
For $U$=19 and 24~{m~s$^{-1}$},
 $P_\infty(p)$  is bimodal with symmetric  peaks, while  $P_\infty(A)$  has its peak shifted away from zero,
which is typical of limit-cycle oscillations (see also  the joint PDF of the acoustic pressure and its time derivative, $P_\infty(p,\dot p/\omega_a)$).
In contrast,
for $U$=17~{m~s$^{-1}$},
$P_\infty(p)$  is unimodal with a peak centered around zero and $P_\infty(A)$ has its peak close to zero.

As shown in Fig.~\ref{fig:prms_wa_spectra}(a), whistling occurs in a tongue-shaped region in the $U$-$\theta$ plane: 
acoustic pressure fluctuations inside the bottle are stronger for larger velocities in the investigated range,
$U \geq 18$~m~s$^{-1}$, 
and intermediate angles, $10^o \leq \theta \leq 50^o$,
reaching up to $p_\text{rms} = 1.4$~mbar (sound pressure level 135~dB).
Above the onset of whistling at a critical velocity $U_c$, acoustic pressure fluctuations increase like $p_\text{rms} \propto \sqrt{U-U_c}$ (Fig.~\ref{fig:prms_wa_spectra}(b)),
in agreement with previous experimental observations \cite{Verge1997a, Dequand2003a}.
To allow comparison with other studies, the inset shows the same data in dimensionless form: amplitude of the vertical acoustic velocity oscillation $v_{\mathrm{ac,rms}}$ in the bottle neck normalized by the jet velocity at the upstream rim $U_j(x_u)$, as a function of the Strouhal number $\Sr = f_a D / U_j(x_u)$ built on the aeroacoustic frequency, neck diameter and local jet velocity.
Specifically, we use mass conservation and the linearized equation of state to deduce the acoustic velocity from our acoustic pressure measurements:
\begin{align} 
v_{\mathrm{ac,rms}} = 
\dfrac{ V \omega_a p_{\mathrm{rms}} }{ \rho c^2 S },
\end{align}
with
$V=3.3 \times 10^{-4}$~m$^3$ the bottle inner volume,
$c = 340$~m~s$^{-1}$ the speed of sound at ambient temperature,
and 
$S=\pi D^2/4 = 2.3 \times 10^{-4}$~m$^2$ the neck cross-section area.
In both representations of Fig.~\ref{fig:prms_wa_spectra}(b), it appears that the acoustic amplitude is still increasing at the largest jet velocity (smallest Strouhal number) investigated. 
From the data of \cite{Dequand2003a} for a similar geometry and a similar ratio of neck diameter to jet width, one can expect the maximum amplitude to be reached at approximately 
$p_{\mathrm{rms}} \simeq 6.5$~mbar and 
$U \simeq 45$~m~s$^{-1}$
(i.e. $v_{\mathrm{ac,rms}}/U_j(x_u) \simeq 0.5$ and 
$\Sr \simeq 0.2$).
One can also note that the critical Strouhal number below which whistling is observed is about 0.5, which is rather consistent with the data in Figs. 8-9 of \cite{Dequand2003a}.

Over the whole range of considered velocities and angles, acoustic pressure signals are almost harmonic and power spectra exhibit a dominant peak at 
$f_a \simeq 195-210$~Hz, as shown in Fig.~\ref{fig:prms_wa_spectra}(c)-(d) 
(where the sound pressure level
is defined as $20\log_{10} \left( p_{\mathrm{rms}} / p_{\mathrm{ref}} \right)$, with the reference sound pressure is $p_{\mathrm{ref}} = 2 \times 10^{-5}$~Pa).
This is consistent with the natural frequency 
$f_0 = c \sqrt{S/(V L_{\mathrm{eq}}) } /(2\pi) \,\simeq 208$~Hz  in air 
 of a Helmholtz resonator of 
equivalent neck length 
$L_{\mathrm{eq}} = 47$~mm.

In the investigated $U-\theta$ range, whistling only occurs for the Helmholtz mode around 210~Hz. Higher modes (the first one around 1550~Hz being visible in Fig.~\ref{fig:prms_wa_spectra}(d)) exhibit negligible acoustic levels in the power spectral density of the acoustic pressure at the bottle bottom.

We note that even at the highest acoustic levels investigated here,
pressure signals remain essentially harmonic, with the peaks of the second and third harmonics two orders of magnitude smaller than that of the fundamental.
Accordingly, the joint PDF $P_\infty(p,\dot p/\omega_a)$ is nearly circular, typical of the phase-space trajectories of weakly non-linear oscillators.

\section{Minimal physical model}
\label{sec:physical_model}

\begin{figure}
\psfrag{f}[t][]{\small $f$ (Hz)}
\psfrag{|H|**2}[b][]{\small $|H|^2$ (-)}
\centerline{   
   \begin{overpic}[height=5cm,tics=10]{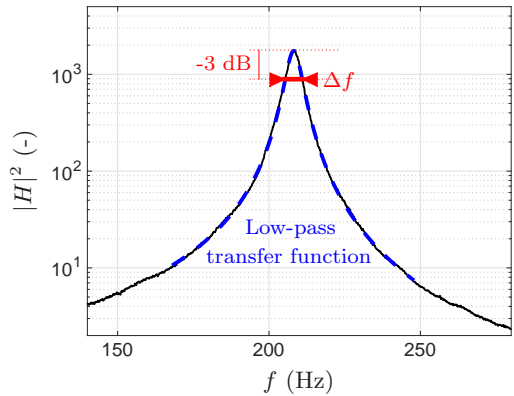}  
      \put(35,65){\footnotesize \tcr{-3 dB}}         
      \put(61,61.5){\footnotesize \tcr{$\Delta f$}}           
      \put(45,31){\footnotesize \tcb{Low-pass}}   
      \put(37,25){\footnotesize \tcb{transfer function}}   
   \end{overpic}  
}
\caption{
Acoustic damping measurement with an external loudspeaker and no jet flow from the pipe.
The bottle's $outside \rightarrow inside$ acoustic transfer function $H$ is reconstructed with cross-PSDs of time signals measured outside and inside the bottle during a series of slow frequency sweeps.
The peak at $f_p=208$~Hz has a quality factor measured directly as $Q=37$ (red, ratio $f_p/\Delta f$ of peak frequency to width at half maximum), which yields an acoustic damping $\alpha=2\pi f_p / Q =35$~rad~s$^{-1}$.
A fit of $|H|^2$ with the squared gain of a second-order low-pass  transfer function (blue) yields a damping value $\alpha=36$~rad~s$^{-1}$.
}  
\label{fig:Q}
\end{figure}

The Helmholtz resonator is a linear acoustic oscillator: in the absence of external forcing, the dynamics of the acoustic pressure $p(t)$ inside the bottle are governed by
\begin{equation}
\ddot p  + \alpha \dot p + \omega_0^2 p
= 0,
\label{eq:linear_osc_acoustic_only}
\end{equation}
where $\alpha>0$ is the acoustic damping and
$\omega_0=2\pi f_0$ is the oscillator's natural angular frequency.
These two quantities are easily obtained experimentally, for instance by measuring the  acoustic transfer function $H(\omega)$ from the outside to the inside of the bottle. 
Figure~\ref{fig:Q} shows the  squared gain of  $H$ measured with a series of frequency sweeps performed with a loudspeaker located outside the bottle.
The value  $\alpha \simeq 35$~rad~s$^{-1}$
is obtained for the acoustic damping, both when calculating  the quality factor (dimensionless ratio $Q=\omega_p/\Delta\omega$ of the peak frequency to the full width at half maximum) and when fitting second-order low-pass or band-pass transfer functions.
The oscillator's natural frequency $\omega_0/(2\pi) = 1320/(2\pi) = 210$~Hz and peak frequency 
$\sqrt{\omega_0^2-\alpha^2/4}/(2\pi)$ 
are almost equal  since damping is weak ($\alpha \ll \omega_0$).

The effect of the grazing jet is twofold: 
a  deterministic forcing induced by coherent  - i.e. acoustically induced -  fluctuations of the advected vorticity,
and a stochastic forcing induced by turbulence.
To illustrate the stochastic identification method with this aeroacoustic system, 
we use a \textit{phenomenological} low-order model for the observable $p(t)$, namely a white-noise-driven VdP oscillator:  
\begin{equation}
\ddot p  + \omega_0^2 p
= (2\nu -\kappa p^2 ) \dot p + \xi(t).
\label{eq:VdP}
\end{equation}
Here
$\nu = (\beta-\alpha)/2$ 
is the linear growth rate
resulting from the combination of the acoustic damping  $\alpha$ and the constructive/destructive aeroacoustic feedback between the jet and the bottle, 
whose linear term is of coefficient $\beta$ and whose saturating non-linear term is of coefficient $\kappa>0$.
The stochastic forcing is modeled as a white Gaussian noise $\xi(t)$ of intensity $\NoiseInt$, i.e. of autocorrelation $\langle \xi\xi_\tau\rangle = \NoiseInt \delta(\tau)$ with $\delta(\tau)$ the Dirac delta function.
The model in Eq.~(\ref{eq:VdP}) is the simplest possible that leads to a limit cycle when the system is linearly unstable ($\nu>0$), and is motivated by the observations made in Sec~\ref{sec:Observations} matching in all respects the characteristics of weakly non-linear self-oscillations.
We wish to stress that this model is not meant to be \textit{predictive}, but to describe \textit{phenomenologically} the observed stochastic bifurcation and dynamics.
More sophisticated analytical descriptions of the aeroacoustic system exist, featuring for instance time delays,
quadratic non-linearities, or colored noise (the interested reader is referred for instance to \cite{Verge1997b, Auvray2012, Auvray2014}). 
In particular, a time delay could easily be added to account for convection effects, which would be consistent with experimental observations of an upper critical jet velocity for whistling. 
Without such an explicit time delay in the model, the parameter identification  returns a good estimate of the actual growth rate, provided that the actual delay does not exceed a couple of acoustic periods.
Therefore, the low-order model in Eq.~(\ref{eq:VdP}) is minimal in the sense that it contains a small number of effective parameters, that can be quantified by analyzing the acoustic time series.

The envelope $A(t)$ and phase $\varphi(t)$ 
are slowly varying compared to the acoustic period $2\pi/\omega_0$, so we write 
$p(t)=A(t) \cos(\omega_0 t + \varphi(t))$ and use deterministic averaging \cite{Krylov1947} and stochastic averaging \cite{Stratonovich1967} to reduce Eq.~(\ref{eq:VdP}) to a system of equations for the amplitude dynamics and phase dynamics
\begin{align}
\displaystyle \dot A = \nu A - \frac{\kappa}{8} A^3 + \frac{\NoiseInt}{4 \omega_0^2 A} + \zeta(t),
\qquad
\dot \varphi = \frac{1}{A} \chi(t), 
\label{eq:Langevin-A}
\end{align}
with $\zeta(t)$ and $\chi(t)$ two independent white Gaussian noises of intensity $\NoiseInt/2\omega_0^2$,
i.e. of autocorrelation 
 $\langle \zeta\zeta_\tau\rangle 
= \langle \chi \chi_\tau \rangle 
= \NoiseInt \delta(\tau) /(2\omega_0^2)$. 
In Eq.~(\ref{eq:Langevin-A}), the Langevin equation for $A$ is independent of $\varphi$ and reads in potential form
\begin{align} 
\dot{A} 
= -\frac{ \mathrm{d}\mathcal{U}}{\mathrm{d}A} + \zeta(t),
\qquad 
\text{with} 
\qquad
\mathcal{U}(A) 
= -\frac{\nu}{2} A^2+ \frac{\kappa}{32} A^4-\frac{\NoiseInt}{4\omega_0^2}\ln(A).
\end{align}
The evolution of $P(A,t)$ is governed by the Fokker-Planck equation 
\begin{equation}
\frac{\partial P}{\partial t}
= -\frac{\partial}{\partial A}  \left(  D^{(1)} P \right)  + \frac{\partial^2}{\partial A^2} \left(  D^{(2)}  P \right),
\label{eq:FPE}
\end{equation}
with drift and diffusion coefficients $D^{(1)}$, $D^{(2)}$ (first two terms of the Kramers-Moyal expansion \cite{Risken84,Stratonovich1967}) 
\begin{equation}
D^{(1)}(A) 
= -\frac{ \mathrm{d}\mathcal{U}}{\mathrm{d}A}
= \nu A - \frac{\kappa}{8} A^3 + \frac{\NoiseInt}{4 \omega_0^2 A},
\qquad 
D^{(2)} = \frac{\NoiseInt}{4\omega_0^2}.
\label{eq:D1D2}
\end{equation}
The stationary PDF $P_\infty(A) = \lim_{t \rightarrow \infty} P(A,t) $ is directly determined by the Kramers-Moyal coefficients:
\begin{align}
P_\infty(A) 
 = \mathcal{N} \exp \left( \frac{\int D^{(1)}(A)}{D^{(2)}}  \right)
= \mathcal{N} \exp \left( - \frac{\mathcal{U}(A)}{\NoiseInt / (4\omega_0^2) }
\right),
\label{eq:PDF_stat}
\end{align}
with $\mathcal{N}$ a normalization factor such that $\int_0^\infty P_\infty(A) \,\mathrm{d}A = 1$.
Note that the mode $A_{m}$ (most probable amplitude, where $P_\infty(A)$ is maximum,
$\mathcal{U}(A)$ is minimum
 and $D^{(1)}(A)=0$) 
differs from the deterministic amplitude $A_{det}=\sqrt{8\nu/\kappa}$.

\section{Parameter identification}
\label{sec:id}

\subsection{Estimation of the Kramers-Moyal coefficients}
\label{sec:KMcoeffs}

We now proceed with the identification of the system's governing parameters $\nu$, $\kappa$, $\NoiseInt$, solely from measured signals of acoustic pressure under stationary conditions.
As mentioned in the introduction,
input-output identification is not possible when the input cannot be measured or when the system cannot be driven arbitrarily.
Noting that the system is driven by a stochastic forcing, we choose as an output-only identification method the estimation of the KM coefficients.
Here we use a model-based version of the method where robustness and accuracy are improved by minimizing the difference between finite-time KM coefficients calculated from time signals
and those calculated with the adjoint FPE \cite{LadePLA09, Honisch11, Boujo2017}.

Specifically, the KM coefficients can be obtained by processing the stationary signal $A(t)$ and extrapolating the first two finite-time  coefficients of the KM expansion to their limit at vanishing time,
\begin{equation}
D^{(n)}(A) = \lim_{\tau \to 0} D^{(n)}_\tau(A),
\quad
D^{(n)}_\tau(A) = \frac{1}{n! \tau} \int_0^\infty (a-A)^n \, P(a, t+\tau | A, t) \, \mathrm{d}a,
\quad n=1,2,
\label{eq:KMcoeff_limit}
\end{equation}
where the integral corresponds to the $n$-th moment of the conditional PDF $P(a, t+\tau | A, t)$, that gives the probability of the amplitude being $a$ at the shifted time $t+\tau$ knowing that it is $A$ at time $t$.
We used 30~s stationary time traces for each  condition $(U,\theta)$.

Alternatively, the finite-time KM coefficients can  be obtained by solving the adjoint FPE for $P^\dag(a,t)$,
\begin{equation}
\frac{\partial P^\dag}{\partial t}
= D^{(1)}  \frac{\partial P^\dag}{\partial a}  + D^{(2)}  \frac{\partial^2 P^\dag}{\partial a^2},
\label{eq:AFPE}
\end{equation}
with a suitable initial condition:
\begin{equation}
P^\dag(a,0) = (a-A)^n 
\quad\Rightarrow\quad
D_\tau^{(n)}(A) = \frac{1}{n! \tau} P^\dag(A,\tau).
\end{equation}

\begin{figure}
\psfrag{U [m/s]}[t][]{\footnotesize $U$ (m~s$^{-1}$)}
\psfrag{nu}[b][]{}
\psfrag{kappa}[b][]{}
\psfrag{Gamma/4omega2}[b][]{}
\psfrag{theta}[r][][1][0]{\footnotesize $\theta$ ($^o$)}
\psfrag{th}[r][][1][-90]{}
\centerline{  
   \begin{overpic}[trim=0 0 0 0, clip, height=5cm,tics=10]{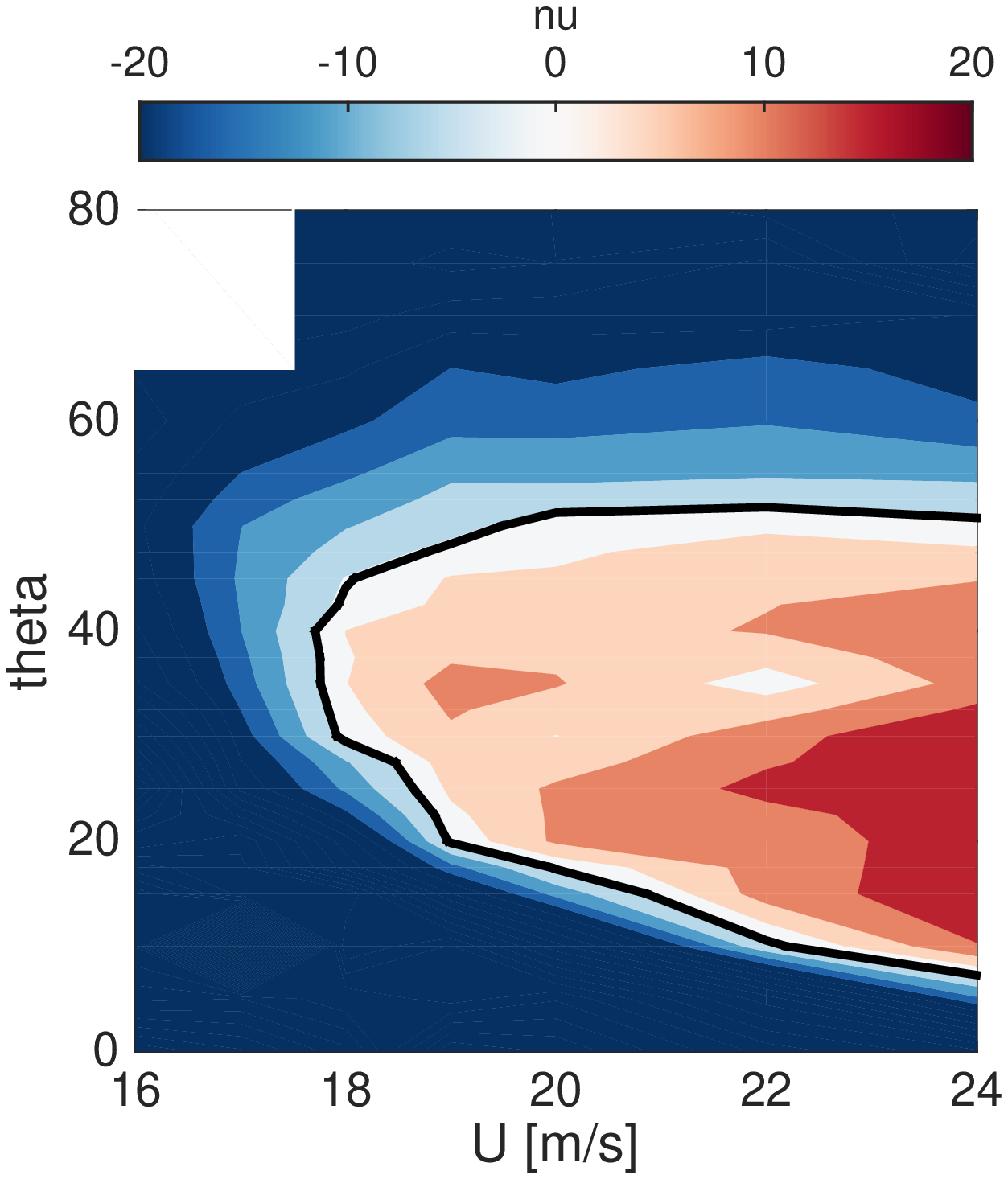}  
      \put(16,73){$\nu$} 
   \end{overpic}  
   \begin{overpic}[trim=0 0 0 0, clip, height=5cm,tics=10]{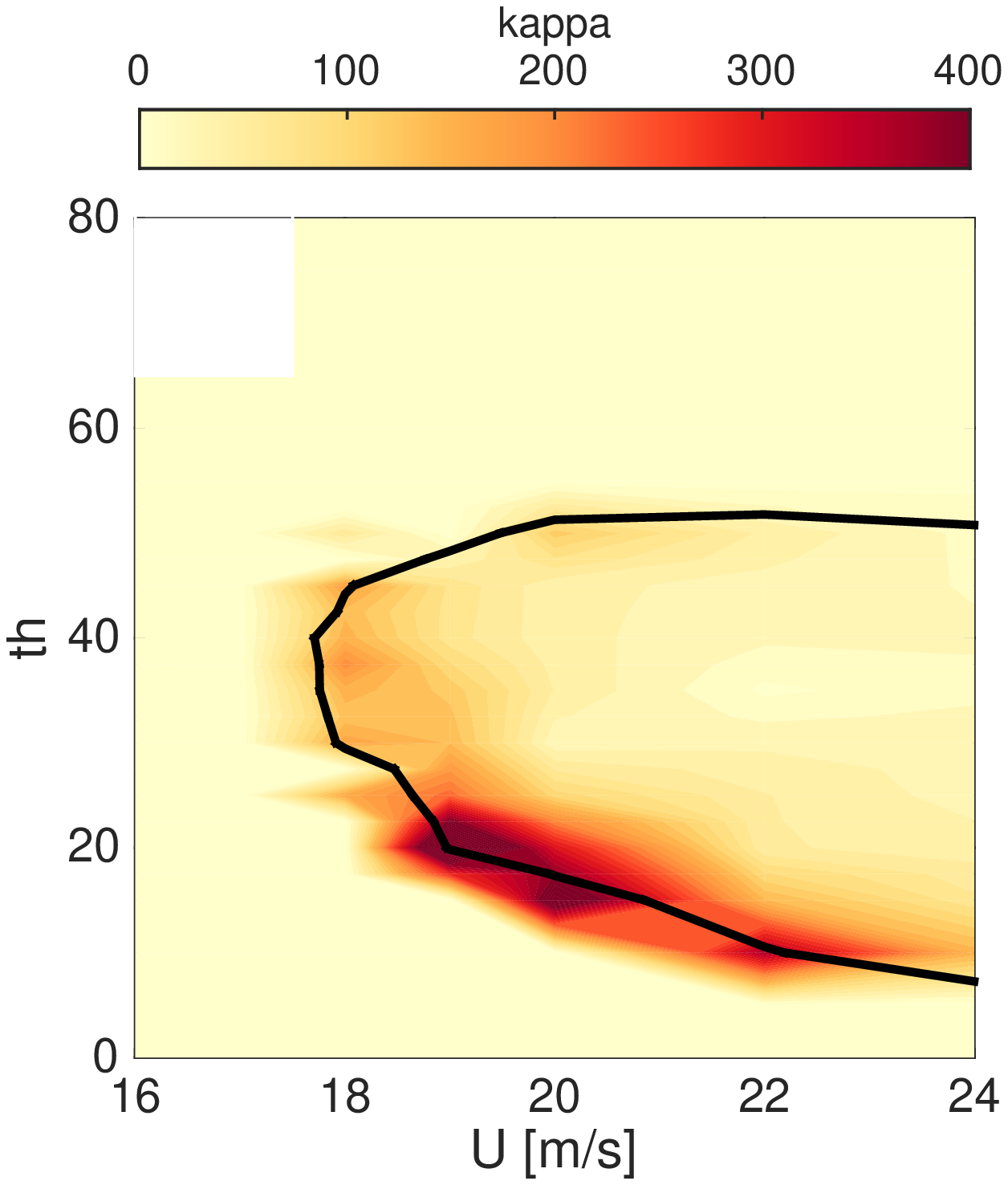}     
      \put(16,73){$\kappa$} 
   \end{overpic}  
   \begin{overpic}[trim=0 0 0 0, clip, height=5cm,tics=10]{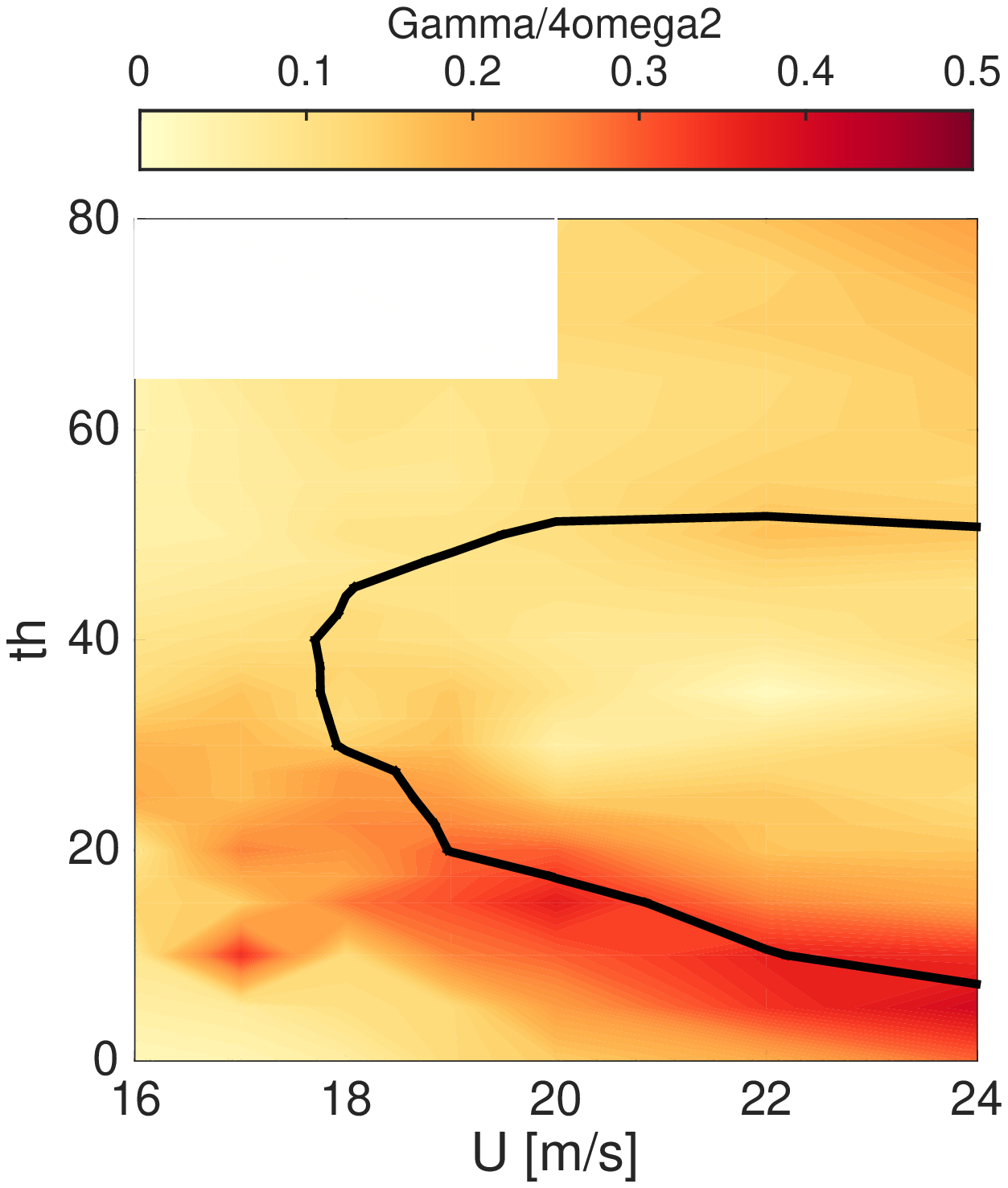}     
      \put(19,73){$\NoiseInt/4\omega_a^2$} 
   \end{overpic} 
}
\caption{
Parameters of the stochastic VdP oscillator Eq.~(\ref{eq:VdP}) and stochastic amplitude equation Eq.~(\ref{eq:Langevin-A}), identified with an adjoint-based optimization method.
Black line: stability limit $\nu=0$.
(Units: $\nu$ in s$^{-1}$, 
$\kappa$ in mbar$^{-2}$s$^{-1}$, 
$\NoiseInt/4\omega_a^2$ in mbar$^{2}$s$^{-1}$.)
}  
\label{fig:coeffs}
\end{figure}

The identification results (Fig.~\ref{fig:coeffs}) show that the oscillator is linearly unstable ($\nu>0$) for larger velocities and intermediate angles, and linearly stable otherwise, with the stability boundary $\nu=0$ following closely the contour
 $p_\text{rms} \simeq 0.4$ mbar
 of the tongue-shaped region in Fig.~\ref{fig:prms_wa_spectra}(a).
$\kappa$ and $\NoiseInt$ are maximum along the stability boundary and for smaller angles. 
This significant asymmetry strikingly contrasts with the rms map (Fig.~\ref{fig:prms_wa_spectra}(a)).
Uncovering the physical mechanisms responsible for the dependency of $\nu$, $\kappa$, $\NoiseInt$ on $U$ and $\theta$ will require further investigation with numerical simulations and PIV measurements of the unsteady flow.

\subsection{Controlled transient experiments: validation of the  parameter identification}
\label{sec:transient_exp}

The output-only parameter identification method is well-suited  for stochastic systems that cannot be controlled; here we take advantage of the fact that the aeroacoustic system can be  controlled with an acoustic forcing, and we  validate quantitatively the identification results. 
In this second set of experiments,  we add an external loudspeaker 20~cm away from the bottle neck, and proceed as follows.

\begin{figure}
\centerline{   
  \begin{overpic}[trim=0 5mm 0 0, clip=true, height=8cm,tics=10]{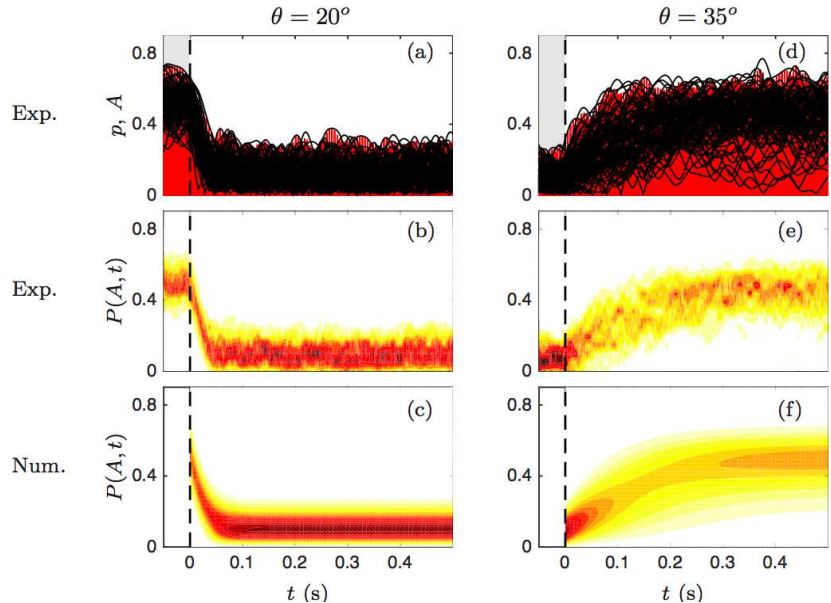}           
  \end{overpic}  
}
\caption{
Transient relaxation dynamics when control from an external loudspeaker is turned off.
(a,d) envelope $A$ (black) of the acoustic pressure $p$ (red) from 100 experimental realizations ($p$ and $A$ in mbar).
(b,e): evolution of the probability density (ensemble average of the envelopes in (a,d) at each time instant).
(c,f): evolution of $P(A,t)$ from the FPE solved in time with the experimental PDF at $t=0$ and with the parameters $\nu$, $\kappa$, $\NoiseInt$ from the adjoint-based identification.
Conditions (gray triangles in Fig.~\ref{fig:prms_wa_spectra}(a)):
$U=18$~m~s$^{-1}$, 
(a-c)~$\theta=20^o$ (linearly stable, relaxation to low-amplitude fluctuations), and 
(d-f)~$\theta=35^o$ (linearly unstable, relaxation to a large-amplitude limit cycle).
}  
\label{fig:controlONOFF}
\end{figure}

First, for linearly stable conditions $(U,\theta)$, the loudspeaker imposes a constant-amplitude forcing at frequency $f_a$ ($t<0$ in Fig.~\ref{fig:controlONOFF}(a)).
At $t=0$, the forcing is switched off and the system relaxes  to its uncontrolled natural state: a stochastically driven linear oscillator. 
Second, for linearly unstable conditions $(U,\theta)$, a feedback control is applied to suppress the limit cycle ($t<0$ in Fig.~\ref{fig:controlONOFF}(d)), 
based on a real-time controller (NI cRIO-9066) coded to delay and amplify the acoustic pressure signal, and to feed the loudspeaker. 
By adjusting the time delay, it is possible to suppress the large-amplitude aeroacoustic limit-cycle. 
At $t=0$, the control is switched off and the system is free to relax to its stable stochastically forced limit cycle. 
In each case, we repeat 100 independent realizations
(Fig.~\ref{fig:controlONOFF}(a), (d)), and compute the ensemble-averaged evolution in the forced/controlled stationary regime and unforced/uncontrolled  transient regime 
(Fig.~\ref{fig:controlONOFF}(b), (e)).

Next, we solve numerically in time the FPE given in Eq.~(\ref{eq:FPE})  (see numerical method in \cite{Boujo2017}), starting from $P(A,0)$ experimentally measured at $t=0$, and using the values of $\nu$, $\kappa$, $\NoiseInt$ from the adjoint-based identification 
(Fig.~\ref{fig:controlONOFF}(c), (f)). 
The time evolution of $P(A,t)$ obtained numerically is 
closely aligned with its experimental counterpart, which validates the parameter identification 
and supports the choice of the low-order stochastic VdP model.
Note that for conditions $U-\theta$ corresponding to linearly stable regimes (for instance those displayed in Fig.~\ref{fig:controlONOFF}(a-c)), the use of the adjoint FPE to identify the decay rate and the noise intensity is rather convoluted, because a simple fit of the Helmholtz resonance peak would suffice. The real strength of this FPE-based methodology comes from its applicability to limit cycle data that are governed by the system non-linearities.

\subsection{Transient simulations: similar statistic, different dynamics}
\label{sec:stat_vs_dyn}

Sections~\ref{sec:KMcoeffs}-\ref{sec:transient_exp} have shown that different stationary \textit{statistics} 
of the stochastic acoustic oscillator correspond to different governing parameters.
We now turn our attention to the effect of these parameters on the system's \textit{dynamic} behavior.

\begin{figure}
\psfrag{t}[r][][-1][90]{\small $t$ (s)}
\psfrag{P}[r][][-1][90]{\small $P_\infty(A)$}
\psfrag{V}[r][][-1][90]{\small $\mathcal{U}(A)$}
\psfrag{A}[t][]{\small $A$ (mbar)}
\psfrag{tau}[][]{ }
\psfrag{<AA>}[][]{ }
\psfrag{A0+/-sig}[][]{ }
\psfrag{G/8}[][]{ }
\vspace{0.05cm}
\centerline{
   \hspace{0.2cm}    
   \begin{overpic}[trim=0mm 5mm 0mm 0,clip=true,width=6cm,tics=10]{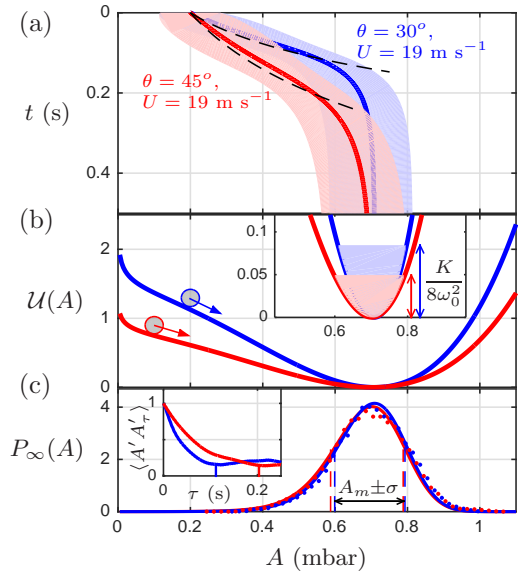}   
      \put(55,94){\scriptsize \tcb{$\theta=30^o$,}} 
      \put(55,90){\scriptsize \tcb{$U=19$~m~s$^{-1}$}} 
      \put(15,84.5){\scriptsize  \tcr{$\theta=45^o$,}}
      \put(15,80.5){\scriptsize  \tcr{$U=19$~\tcr{m~s$^{-1}$}}}
      \put(-8,95){(a)}
      \put(-8,58){(b)}
      \put(-8,27){(c)}
      \put(67.5,47){\scriptsize $\dfrac{\NoiseInt}{8\omega_0^2}$}
      \put(51.5,8){\scriptsize $A_{m}$}
      \put(57.5,8){\scriptsize $\pm$}
      \put(60.5,8){\scriptsize $\sigma$}
      \put(23.5,7){\scriptsize $\tau$ (s)}
      \put(12.5,12.1){\rotatebox{90}{\scriptsize $\langle A' A'_\tau \rangle$}}
   \end{overpic}   
}
\vspace{0.2cm}
\caption{
(a)~Time evolution
$P(A,t)$ for two conditions (circles in Fig.~\ref{fig:prms_wa_spectra}(a)) exhibiting different transient dynamics and yet leading to the same stationary statistic.
Numerical simulation of the FPE with the identified $\nu$, $\kappa$, $\NoiseInt$ from the adjoint-based identification.
Colored areas: $P(A,t)$ larger than a fixed given value.
The most probable amplitude $A_{m}(t)$ (thick lines) compares well with the exponential growth $A_{m}(0) \mathrm{e}^{\nu t}$ (dashes).
(b)~Identified potential $\mathcal{U}(A)$ (in mbar$^2$s$^{-1}$).
Inset: detail of the  well around $A_{m}$.
Colored areas: typical potential height $\NoiseInt/(8\omega_0^2)$ (in mbar$^2$s$^{-1}$) reached when the system visits amplitudes in the range $A_{m} \pm \sigma$ (one standard deviation away from the most probable amplitude).
(c)~$P_\infty(A)$: experimental measurements (symbols) and analytical expression with the identified parameters (solid lines).
Inset: normalized autocorrelation (dimensionless) of the fluctuations $A' = A - \overline A$,
whose first minimum is at the characteristic time $\tau \simeq 1/\nu$.
}
\label{fig:same_pdf_different_dynamics}
\end{figure}

One can note that different sets of parameters may result in similar acoustic levels  (Figs.~\ref{fig:prms_wa_spectra}(a) and \ref{fig:coeffs}) and similar PDFs.
For instance, for $U=19$~m~s$^{-1}$, both jet angles
$\theta=30^o$ and 45$^o$ lead to $p_\text{rms}=0.50$~mbar (red and blue circles in Fig.~\ref{fig:prms_wa_spectra}(a))
and to similar stationary PDFs 
(Fig.~\ref{fig:same_pdf_different_dynamics}(c), 
showing a good agreement between measurements and identification), 
while system identification yields 
radically different sets of parameters:
$\{\nu,\kappa,\NoiseInt/4\omega_a^2\} 
=\{9.0,\,148,\,0.17\}$ and 
$\{4.9,\,83,\,0.10\}$ respectively 
(units: s$^{-1}$,  mbar$^{-2}$s$^{-1}$,  and mbar$^{2}$s$^{-1}$ respectively).
This can be explained as follows. 
Compared to  $\theta=45^o$, the operating condition
$\theta=30^o$ is characterized by a larger growth rate $\nu$ and a stronger saturation $\kappa$, leading to a well of the potential $\mathcal{U}(A)$ that is steeper, as shown in Fig.~\ref{fig:same_pdf_different_dynamics}(b).
In parallel, the identified noise intensity $\NoiseInt$ is higher too, so the system is able to explore higher regions of $\mathcal{U}(A)$, as illustrated by the typical potential height $\NoiseInt/(8\omega_0^2)$ reached when the amplitude departs by one standard deviation from its most probable value, $A=A_m\pm\sigma$. 
The net result is that the two systems have 
the same stationary \textit{statistic}.

However, their \textit{dynamics} differ. 
Figure~\ref{fig:same_pdf_different_dynamics}(a) shows the evolution of $P(A,t)$ when starting from a non-equilibrium PDF centered around low amplitudes. 
Both systems relax to the same stationary  $P_\infty(A)$ but at 
different rates,
because the underlying potentials 
have different depths.
Note that those relaxation rates are consistent with the identified linear growth rates $\nu$, as shown by the dashed lines $A_{m}(0) \mathrm{e}^{\nu t}$.)
{This difference in dynamics is also observed} in the autocorrelation of the fluctuations of $A(t)$, which shows different characteristic times scales $\tau \simeq 1/\nu$ for the two systems (inset in Fig.~\ref{fig:same_pdf_different_dynamics}(c)). 
Therefore, for any stationary condition, 
$U$ and $\theta$ have a strong influence on the dynamics of the system on its way to the limit cycle. 
This can have a significant influence on transient regimes, e.g. in music (attack transients or changes between different regimes) \cite{Auvray2012, Auvray2014},
and in engineering applications \cite{Bonciolini2018, Bonciolini2019a, Bonciolini2019b}.

\section{Discussion}
\label{sec:discussion}

Some comments are in order about the practical use of the present output-only identification method for stochastically driven oscillators.
We start in Sec.~\ref{sec:discuss_KM} with comments related to the estimation of the Kramer-Moyals coefficients, because as already mentioned, this can be done in general without an \textit{a priori} model of the system \cite{Boettcher06,Friedrich2011}.
Here, accuracy is improved thanks to the adjoint FPE, which requires a model, i.e. an explicit expression of the KM coefficients such as Eq.~(\ref{eq:D1D2}), so we continue in Sec.~\ref{sec:discuss_VdP} with comments related to the specific model used in this study.

\subsection{Estimation of the KM coefficients}
\label{sec:discuss_KM}

Independently of any model, directly computing the drift and diffusion coefficients from the Hilbert transform of a quasi-periodic observable
allows us to disentangle deterministic and stochastic components of the slow-flow components (amplitude and phase) of the system subject to dynamic noise. 
Some of the associated limitations are listed below.
\begin{itemize}
\item
\textit{Signal length:}
time traces must be long enough to obtain converged  statistics of the envelope.
The required signal duration is linked to the linear growth rate to be identified.
\item
\textit{Stationarity:}
it is important that the time traces exhibit the characteristics of a stationary process, which is a key assumption for the present analysis of the extracted KM coefficients. In particular, there should be no significant linear trend.
On the other hand, rare events are allowed.
\item
\textit{Dominant dynamics:}
if, in contrast to the present experiment, the self-oscillation does not dominate the dynamics because other modes of the system display non-negligible contributions in the power spectral density of the time trace,
the processing has to be performed after band-pass filtering the data around the frequency of the mode of interest, in order to isolate its specific dynamics and statistic.
\item
\textit{Quasi-harmonicity and slow dynamics:}
processing oscillation amplitude data to infer properties of an oscillator subject to random  forcing only works under the hypothesis that the system exhibits weakly non-linear self-oscillations. 
When this is not the case, one cannot properly define a slowly varying envelope, but  
identification may still be possible using acoustic pressure signals.
\item
\textit{Noise intensity:}
the intensity of the dynamic noise should be large enough such that the system is randomly forced to reach states that are significantly away from the deterministic attractor. 
Consequently, the observable contains information that allows us with this methodology to properly characterize a large region of the manifold that defines the slow-flow dynamics, and get hints about the ingredients of a minimal low-order model. In other words, one of the main limitations of the method is that it does not work for quasi-deterministic systems. 
\item
\textit{Finite-time effects:}
evaluating the limit in Eq.~(\ref{eq:KMcoeff_limit}) involves an extrapolation to vanishing time shift, which may be inaccurate due to finite-time effects such as coarse sampling rate, 
band-pass filtering of the data, 
or Markov property not holding because of a dynamic noise not strictly delta-correlated. 
Using the adjoint Fokker-Planck equation addresses exactly this limitation \cite{LadePLA09, Honisch11, Boujo2017}.
\end{itemize}

\subsection{Specific model}
\label{sec:discuss_VdP}

The above data processing (extraction of the KM coefficients) guides us in guessing a minimal model  such that the linear growth rate of the system can be identified from limit-cycle data. 
In the present case, performing the simple KM analysis and extrapolating finite-time moments to zero time shift yields drift and diffusion coefficients that are compatible with the Langevin equation for the amplitude of a simple Van der Pol oscillator, and this over the range of operating conditions investigated.  
In addition, different other hints bring us to the conclusion that the VdP model is an appropriate minimal model: 
clear supercritical Hopf bifurcation defining the stability border for a range of combinations $\theta-U$, 
square root increase of the amplitude with the jet velocity, 
PDF of the acoustic pressure, 
PDF of the slowly varying envelope and phase, 
low-pass behavior of the envelope. 
This was also the case when identifying with this approach the linear growth rate of the thermoacoustic instability investigated in \cite{NoirayDenisov16},  which constitutes another example of application.

However, in the present case, some observations also show that there are ingredients missing in our low-order model, as discussed below. 
\begin{itemize}
\item
\textit{Non-linearity:}
Harmonics at third, but also at twice the fundamental frequency in Fig.~\ref{fig:prms_wa_spectra}
suggest that there is a non-linearity of the aeroacoustic feedback leading to a transfer of energy from the self-sustained oscillation at the fundamental frequency to oscillations at twice the frequency. 
The cubic term in the VdP model must not be interpreted as a complete description of the non-linear dynamics, but as an effective description that reproduces the saturation at fundamental frequency with a redistribution of energy at the third harmonic only, while in reality it happens at all the harmonics. 
 The energetic contribution of the harmonics to the limit cycle is orders of magnitude lower than at the fundamental, therefore, since we focus on a quantitative identification of the linear growth rate only, it does not matter if the non-linear description is incomplete, provided that our minimal model satisfactorily reproduces the shape of the potential well (from the envelope PDF and its left and right tails, it looks very satisfactory). 
For an effective description of the non-linear dynamics significantly away from the supercritical Hopf bifurcation point, 
other non-linear terms may be needed.
 Also, this effective description with solely a cubic term is only suited for oscillators exhibiting a supercritical Hopf bifurcation. For the case of non-linearities leading to subcritical Hopf
 bifurcations, e.g. \cite{Bonciolini2018}, the model should also include quadratic and quintic terms for an effective description of the non-linearities. 
\item 
\textit{Colored noise:}
Our model is for sure incomplete with regard to the nature of the additive stochastic forcing. 
With the same argumentation as before, the white Gaussian noise (an ideal representation, never truly satisfied in practice) in our minimal model must not be interpreted as a complete description of the actual forcing from turbulence. 
The energy content of the turbulent forcing is broadband, and assuming that it is uniformly distributed over a narrow frequency range around the fundamental frequency is a plausible assumption. 
Therefore, since we focus on a quantitative identification of the linear growth rate only (we do not attempt to identify the spectral distribution of the broadband noise), we can work with our minimal model (also discussed in a thermoacoustic context in \cite{Bonciolini2017}). 
\item
\textit{Number of parameters:}
the reliability of the output-only model-based identification method significantly depends on the number of parameters to identify. This is why it is crucial to first start the analysis with a minimal low-order model and critically think about the possibility to extract more information from the observable. 
For instance, the linear growth rate results from the combination of the feedback force delay, the feedback force amplitude and the acoustic losses, therefore it is impossible to independently characterize these three quantities from the acoustic pressure only;
still, with an additional simultaneously-recorded observable, which could be in the present case a long time trace of the vertical component of the spatially-integrated vortex force, one may aim at identifying the parameters of more complex low-order models with explicit formulation of losses, feedback gain and mean delay. The reader can refer to \cite{Boujo2016chemi} for an example of such methodology extension in the context of thermoacoustic instabilities.    
\item 
\textit{Time delay:}
 there is for sure a time delay in the feedback due to the spatially distributed nature of the fluctuating vortex force involved in the aeroacoustic feedback. 
The spatio-temporal evolution of its vertical component (Fig.~\ref{fig:P}) shows that the feedback is delayed by an effective convective time that is shorter than one acoustic period. 
Including such a time delay could explain the critical velocity and angle for the onset of whistling.
The linear growth rate of the system is defined by the effective delay, the amplitude of the associated spatially integrated force, and the acoustic losses. 
Since only the linear growth rate is of interest to us, it can be identified with our minimal model. 
\end{itemize}

\section{Conclusion}

We apply a model-based, output-only system identification to an experimental aeroacoustic self-sustained oscillator subject to random forcing: a Helmholtz resonator made of a jet impinging the neck of a beer bottle, at different velocities and angles.
The method requires no input, and the measurement of a single output, namely the acoustic pressure inside the bottle. 
Noting that pressure oscillations have the characteristics of a weakly non-linear oscillator subject to dynamic noise, we choose to describe 
the system phenomenologically, in the range of investigated operating conditions, 
with a simple low-order stochastic model, namely a Van der Pol oscillator with additive stochastic forcing.
Importantly, the identification method relies on the presence of noise, which allows a stochastic description with a Langevin equation, a Fokker-Planck equation, and the corresponding Kramers-Moyal coefficients.
Computing the finite-time transition moments of the acoustic pressure envelope, and using an optimization technique based on the adjoint Fokker-Planck equation, we identify the parameters of the VdP potential well (in particular the linear growth rate) and of the stochastic forcing for different operating conditions.
The results of the identification method are validated with controlled experiments.

We observe from the identification results that the method gives insight into the competing phenomena leading or not to a limit cycle.
In addition, we note that different operating conditions may correspond to similar long-term statistics (e.g. mean and fluctuations of the pressure envelope) but to different dynamic properties (e.g. rate of transition to a limit cycle), which is explained in terms of potential shape and noise intensity.
The low-order stochastic model and the model-based identification are useful for disentangling deterministic and stochastic effects in systems that exhibit stochastic dynamics but that cannot be controlled arbitrarily, so we expect further applications in aeroacoustics as well as other fields.

An interesting extension of the  method would be the identification of multi-dimensional observables. 
For instance, (i)~individual contributions to the linear growth rate (e.g. time delay, losses, feedback) could be identified by measuring  several observables simultaneously \cite{Boujo2016chemi};
(ii)~several modes that have nearby frequencies and whose mutual influence cannot be filtered out could be identified simultaneously with a multi-oscillator model;
(iii)~other types of non-linearities (e.g. stiffness non-linearities in the Duffing and Duffing-Van der Pol oscillators) that lead to coupled dynamics for the slowly-varying phase and amplitude could be identified with a two-dimensional Fokker-Planck equation.

\section*{Acknowledgments}
This work was supported by Repower and the ETH Zurich Foundation.

\section*{References}

\bibliography{bottle_arxiv_2.bbl}

\end{document}